\documentclass[emulateapj,apj]{emulateapj}

\slugcomment{Not to appear in Nonlearned J., 45.}

\shorttitle{GRAVITATIONAL FRAGMENTATION OF EXPANDING SHELLS. I.}
\shortauthors{Iwasaki, Inutsuka, \& Tsuribe}

\begin{document}


\title{GRAVITATIONAL FRAGMENTATION OF EXPANDING SHELLS. I. LINEAR ANALYSIS}


\author{Kazunari iwasaki\altaffilmark{1}, Shu-ichiro Inutsuka\altaffilmark{1}, and 
Toru Tsuribe\altaffilmark{2}}

\altaffiltext{1}{Department of Physics, Nagoya University, Furo-cho, 
Chikusa-ku, Nagoya, Aichi, 464-8602, Japan; 
iwasaki@nagoya-u.jp, inutsuka@nagoya-u.jp}
\altaffiltext{2}{Department of Earth and Space Science, Osaka University, Machikaneyama-cho 1-1, Toyonaka, Osaka, 
560-0043, Japan:\\
tsuribe@vega.ess.sci.osaka-u.ac.jp}




\begin{abstract}
We perform a linear perturbation analysis of expanding shells driven 
by expansions of H{\it II} regions. 
The ambient gas is assumed to be uniform.
As an unperturbed state, 
we develop a semi-analytic method for deriving the time evolution of the 
density profile across the thickness.
It is found that the time evolution of the density profile 
can be divided into three evolutionary phases, deceleration-dominated, intermediate,
and self-gravity-dominated phases. 
The density peak moves relatively from the shock front to the contact discontinuity as 
the shell expands.
We perform a linear analysis taking into account the asymmetric density profile 
obtained by the semi-analytic 
method, and imposing the boundary conditions for the shock front and 
the contact discontinuity while
the evolutionary effect of the shell is neglected.
It is found that the growth rate is enhanced compared with the 
previous studies based on the thin-shell approximation.
This is due to the boundary effect of the contact discontinuity and 
asymmetric density profile that were not taken into account in previous works.
\end{abstract}



\section{Introduction}
Expanding shells are ubiquitous in the interstellar medium.
They are driven by energetic phenomena of massive stars, such as 
emission of ionizing photons, stellar winds, and supernova explosions.
Recently, using 102 samples identified as shell, \citet{Detal10}
found evidences of the star formation in more than a quarter of the shells, 
suggesting that the triggered star formation by HII regions  may be 
an efficient process of the massive star formation.
Theoretically, 
\citet{EL77} presented a sequential star formation scenario where 
the massive star formation takes place through gravitational fragmentation 
of the expanding shell 
that is driven by HII regions surrounding massive stars, and 
newly formed massive star also triggers the formation of next generation.

To understand the triggered star formation,
it is important to investigate how and when the expanding shell fragments 
through the gravitational instability (GI).
Earliest studies were done by using linear analyses of the 
static dense gas layer confined 
by the same thermal pressures of hot rarefied gases on both sides
\citep{GL65,EE78,LP93}.
They showed that the GI begins to develop with 
a scale comparable to the layer thickness and with a growing timescale 
comparable to the free-fall time of the layer.
However, their linear analyses are oversimplified because the actual shells are
confined by the shock front (SF) on the leading surface and the contact 
discontinuity (CD), or the ionization front (IF) on the trailing surface. 
Moreover, an unbalance between the ram pressure and the thermal pressure 
causes a decelerating or an accelerating expansion.
Many authors have tackled the stability analyses with these effects 
by mainly using the thin-shell approximation where the perturbed variables are averaged 
across the thickness. 
The stability analysis of expanding shells has been investigated by 
\citet{V83}, \citet{E94}, and \citet{Wetal94b}. 
They took into account dilution effects of perturbations owing 
to the expansion and the mass accretion. 
Their linear analyses of expanding shells 
neglected the structure across the thickness and the boundary effect of the CD. 
Thus, how these effects that they neglected influence the GI have been unknown yet.
\citet{V88} investigated stability of asymmetric layers,
and found that the asymmetry of the density profile of the shell greatly influences the 
development of the GI. 
Moreover, by using shock-like boundary conditions, he found 
that the different choice of the boundary condition greatly
modifies the dispersion relation.
However, their analysis is limited to be in the incompressible fluid.

In this paper, we perform a linear analysis taking into account 
the structure across the thickness 
and the effects of boundaries, i.e., the SF on the leading surface and 
the CD on the trailing surface.
In order to determine the density profile all the time, we 
develop a semi-analytic 
method that well describes the one-dimensional (1D) evolution.
This paper extends the study of \citet{V88} to include the 
compressible effect and the more realistic density profile by 
taking into account the radial self-gravitational force \citep{WF02}.
We neglect the effects of expansion and mass accretion through the SF.

In this paper, since we focus on investigation of
how the boundary effects and asymmetric density profiles
influence the GI, we do not apply our result to estimate fragmentation time and scale. 
We will perform three-dimensional simulation of expanding shells to 
compare with the results of the linear analysis, and  
present detailed quantitative aspects of the 
fragmentation process of expanding shells in a subsequent paper 
\citep[][submitted]{IIT11}.

The outline of the paper is as follows:
in Section \ref{sec:HII}, we present a thin-shell model of the expanding shell 
driven by the HII region.
In Section \ref{sec:unp}, we develop a semi-analytic method to derive 
time evolution of the density profile.
We investigate influences of the asymmetric density profile on the 
dispersion relation of the GI by considering pressure-confined layer 
in Section \ref{sec:asym}.
In Section \ref{sec:shell}, we perform linear analysis of expanding shells 
by using density profile obtained 
in Section \ref{sec:unp} and by imposing the approximate SF 
and the CD boundary conditions. 
In Section \ref{sec:discuss}, we compare our results with previous 
works.
Summary is presented in Section \ref{sec:summary}.

\section{Thin-Shell Model Driven by HII Region}\label{sec:HII}
Massive stars emit ultraviolet photons ($h\nu>13.6$ eV) and produce  
HII regions around them.
Here, we consider a massive star that emits ionizing photons with the photon number 
luminosity 
$Q_\mathrm{UV}\:[\mathrm{s}^{-1}]$, into the ambient gas with the 
uniform density of $\rho_\mathrm{E}=mn_\mathrm{E}$, where 
$n_\mathrm{E}$ and $m$ are the number density and the mean mass of the ambient gas particle, respectively.
In the standard picture \citep[e.g.,][]{S78}, the IF initially expands with 
a supersonic speed with respect to the sound speed of ionized gas, 
$c_\mathrm{II}$.
The HII region 
begins to expand by the pressure difference between the HII region and 
the ambient gas when the IF reaches the Str{\"o}mgren radius, $R_\mathrm{ST}$  given by 
\begin{equation}
        R_\mathrm{ST} = \left( \frac{3Q_\mathrm{UV}}{4\pi \alpha_\mathrm{B} n_\mathrm{E}^2} \right)^{1/3},
\end{equation}
where $\alpha_\mathrm{B}$ indicate the case-B recombination coefficient. 
In this phase, the SF emerges in front of the IF and sweeps up the ambient 
gas into a dense shell.
This paper focuses on the evolution of the shell after the shock emerges.
The equation of motion of the shell is given by
\begin{equation}
        \frac{\mathrm{d} }{\mathrm{d} t}\left(M_\mathrm{s}
 \frac{\mathrm{d} R_\mathrm{s}}{\mathrm{d} t} \right)
 = 4\pi R_\mathrm{s}^2 P_\mathrm{II},
\label{eom shell}
\end{equation}
where $M_\mathrm{s}=4\pi G \rho_\mathrm{E} R_\mathrm{s}^3/3$ is the total mass of the shell, i.e.
the mass of the ambient gas that initially occupied the volume of the HII region, 
$R_\mathrm{s}$ is the mean radius of the shell and $P_\mathrm{II}$ is the thermal 
pressure of the HII region.
Here, we neglect the pressure of the ambient gas and the thickness of the shell.
In the HII region, the detailed balance between the recombination and the ionization 
is approximately established all the time. Therefore,
$P_\mathrm{II}$ can be expressed using $R_\mathrm{s}$ as follows:
\begin{equation}
        P_\mathrm{II} = \rho_\mathrm{E} c_\mathrm{II}^2 
        \left( \frac{R_\mathrm{ST}}{R_\mathrm{s}} \right)^{3/2}.
    \label{Pii}
\end{equation}
Using Equation (\ref{Pii}), we obtain the solution of Equation (\ref{eom shell}), 
\begin{equation}
     R_\mathrm{s}(t) = 
     R_\mathrm{ST}\left( 1+\frac{7}{\sqrt{12}}\frac{c_\mathrm{II} t}{R_\mathrm{ST}} \right)^{4/7} 
     \label{thin-shell}
\end{equation}
\citep{HI06}.
Equations (\ref{eom shell}) and (\ref{thin-shell}) are valid only in the early phase.
As the shell sweeps up the ambient gas and increases its mass, the self-gravity 
influences the expansion.
The equation of motion including the self-gravity becomes
\begin{equation}
        \frac{\mathrm{d} }{\mathrm{d} t}\left( M_\mathrm{s}
        \frac{\mathrm{d} R_\mathrm{s}}{\mathrm{d} t} \right)
        = 4\pi R_\mathrm{s}^2 P_\mathrm{II} 
        - \frac{G M_\mathrm{s}^2}{2R_\mathrm{s}^2},
\label{eom shell grav}
\end{equation}
where the second term on the right-hand side 
represents the self-gravitational force \citep{WF02}.
The factor of $1/2$ in the self-gravity term arises because the 
gravitational acceleration
vanishes at the inner surface, it is $GM_\mathrm{s}/R_\mathrm{s}^2$ 
at the outer surface, and the mass-weighted average across the thickness 
is $GM_\mathrm{s}/2R_\mathrm{s}^2$.
One can see that the self-gravity slows the expansion in 
Equation (\ref{eom shell grav}). 

In this paper, for convenience, the units of the time, length, 
and mass scales are taken to be
\begin{equation}
  t_0 = \sqrt{\frac{3\pi}{32G\rho_\mathrm{E}}} = 
  1.6\;n_\mathrm{E,3}^{-1/2}\;\mathrm{Myr},
  \label{t scale}
\end{equation}
\begin{equation}
  R_0 = \left( \frac{7 c_\mathrm{II} t_0}{\sqrt{12}} \right)^{4/7} R_\mathrm{ST}^{3/7}
  = 5.9\;
  Q_\mathrm{UV,49}^{1/7}\:
  n_\mathrm{E,3}^{-4/7}\;\mathrm{pc},
  \label{r scale}
\end{equation}
and
\begin{equation}
  M_0 = \rho_\mathrm{E} R_0^3
  = 5.0\times10^3\;
  Q_\mathrm{UV,49}^{3/7}\:
  n_\mathrm{E,3}^{-5/7}\;M_\odot,
  \label{m scale}
\end{equation}
respectively, where $Q_\mathrm{UV,49}=Q_\mathrm{UV}/10^{49}\:\mathrm{s}^{-1}$,
and $n_\mathrm{E,3}=n_\mathrm{E}/10^3\:\mathrm{cm^{-3}}$.

Non-dimensional quantities normalized 
by $t_0$, $R_0$, and $M_0$ 
are expressed by using tilde, e.g., $\tilde{R}_\mathrm{s}=R_\mathrm{s}/R_0$.
Using non-dimensional quantities, 
we can rewrite Equations (\ref{Pii}) and (\ref{eom shell grav}) as 
\begin{equation}
        \tilde{P}_\mathrm{II} = \frac{12}{49}\tilde{R}_\mathrm{s}^{-3/2},
        \label{nondim PII}
\end{equation}
and
\begin{equation}
 \frac{\mathrm{d} }{\mathrm{d} \tilde{t}}\left( \tilde{R}_\mathrm{s}^3 \frac{\mathrm{d} \tilde{R}_\mathrm{s}}{\mathrm{d} \tilde{t}} \right)
 = \left(\frac{6}{7}\right)^2\tilde{R}_\mathrm{s}^{1/2} - \frac{\pi^2}{16}\tilde{R}_\mathrm{s}^4,
\label{eom shell nondim}
\end{equation}
respectively.
\begin{figure}[htpb]
     \begin{center}
         \includegraphics[width=8cm]{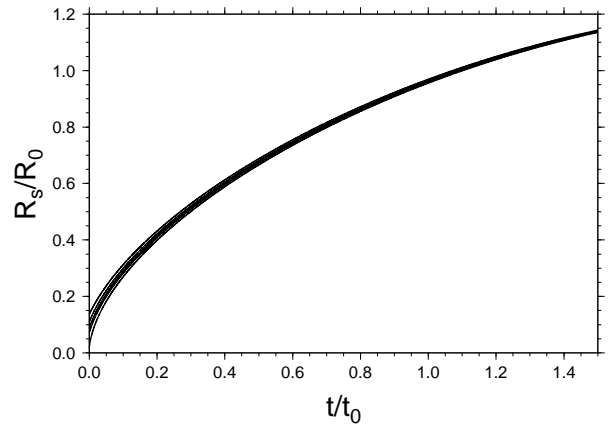}
     \end{center}
\caption{
Expansion laws of shells.  
The abscissa and ordinate axes indicate the time $t/t_0$ and the radius of the shell $R_\mathrm{s}/R_0$, 
respectively.
The solid lines correspond to the case with ($n_\mathrm{E}/\mathrm{cm^{-3}}$, 
$Q_\mathrm{UV}/\mathrm{s^{-1}}$)$=(10^3$, $10^{49})$, $(10^2$, $10^{49})$, 
$(10^4$, $10^{49})$, $(10^3$, $10^{48})$, and $(10^3$, $10^{45})$.
}
\label{fig:Quv n}
\end{figure}
We integrate Equation (\ref{eom shell nondim}) with respect to time with 
the initial condition, 
$\tilde{R}=\tilde{R}_\mathrm{ST}$ at $\tilde{t}=0$.
The initial velocity $\mathrm{d}\tilde{R}_\mathrm{ST}/\mathrm{d}\tilde{t}$ is 
obtained from Equation (\ref{thin-shell}) with $\tilde{t}=0$.
Figure \ref{fig:Quv n} shows 
the obtained expansion law 
with various parameters, $(n_\mathrm{E}/\mathrm{cm^{-3}}$, 
$Q_\mathrm{UV}/\mathrm{s^{-1}})=(10^3$, $10^{49})$, $(10^2$, $10^{49})$, 
$(10^4$, $10^{49})$, $(10^3$, $10^{48})$, and $(10^3$, $10^{45})$.
The difference of these parameter gives the different values of 
$\tilde{R}_\mathrm{s}$ at $\tilde{t}=0$ as shown in Figure \ref{fig:Quv n}.
In Figure \ref{fig:Quv n}, it is seen that 
as the shell expands, the lines quickly approach to an 
asymptotic line that is independent of the parameters.
Therefore, the dependence of the expansion law on the parameters is
approximately eliminated by using the non-dimensional quantities.

\citet{WF02} derived similar thin-shell equations for the shells driven by 
steady stellar winds. They found a change of the power-law index (from 3/5 to 1/5) at the time
when the self-gravity starts to be important. In the gravity dominated phase, the shell expands keeping
the force balance between the thermal pressure of the hot bubble and the self-gravity.
In the stellar wind case, the steady energy input allows the outward expansion of the shell 
($\propto R_\mathrm{s}^{1/5}$) even when the self-gravity becomes important.
On the other hand, in the case with the HII regions, from Equation (\ref{eom shell nondim}),
the gravitational force ($\propto \tilde{R}_\mathrm{s}^4$) 
increases more rapidly than the pressure force by the HII region ($\propto \tilde{R}_\mathrm{s}^{1/2}$),
suggesting that the shell begins to collapse toward the center at a certain radius. In 
the numerical calculation, it occurs at $\tilde{R}_\mathrm{s}\sim2.3$ in all parameters.
The last term of Equation (\ref{eom shell nondim}) is valid until only the expansion phase.
In reality, besides ionizing photon, the massive star emits strong stellar wind continuously over several tens 
of million years and dies through supernova explosion \citep{W77}.
They may influence the dynamics of the shell in the self-gravity dominated phase.
In this paper, for simplicity, we focus on the expansion phase by the ionizing photon.

\section{Time Evolution of Density Profiles: Unperturbed State}\label{sec:unp}
In this section, we derive the time evolution of the density profile of the shell in a 
semi-analytic way.
We assume that the shell is in instantaneous hydrostatic equilibrium at each instant of time.
This is reasonable assumption because the shell is very thin and the sound-crossing time 
across the thickness is very short compared with the expansion timescale.
The equation of the hydrostatic equilibrium in the frame of the shell is given by
\begin{equation}
        -\frac{c_\mathrm{s}^2}{\rho}\frac{\mathrm{d} \rho}{\mathrm{d} r} - \frac{\mathrm{d} \phi}{\mathrm{d} r} 
        + g_\mathrm{dec}=0,
        \label{hydrostatic}
\end{equation}
where $g_\mathrm{dec} = -\mathrm{d}^2 R_\mathrm{s}/\mathrm{d} t^2$ is the inertial force 
owing to the deceleration of the shell, and is assumed to be spatially 
constant within the shell.
In the decelerating shell, the inertia force is parallel to the radial direction.
The Poisson equation is 
\begin{equation}
	\frac{\mathrm{d}^2 \phi}{\mathrm{d} r^2} + \frac{2}{r}\frac{\mathrm{d} \phi}{\mathrm{d} r}\simeq
	\frac{\mathrm{d}^2 \phi}{\mathrm{d} r^2} = 4\pi G \rho,
        \label{poisson}
\end{equation}
where the curvature effect is neglected
because $R_\mathrm{s}$ is much larger than the thickness.
We confirmed that the curvature effect is negligible by comparing density profiles
with and without curvature effect.
Substituting Equation (\ref{hydrostatic}) into Equation (\ref{poisson}),
one obtains
\begin{equation}
        \frac{\mathrm{d} }{\mathrm{d} r}\left( \frac{c_\mathrm{s}^2}{\rho}
        \frac{\mathrm{d} \rho}{\mathrm{d} r} \right) = -4\pi G \rho.
        \label{hydro stat}
\end{equation}
Equation (\ref{hydro stat}) can be solved analytically as follows:
\begin{equation}
        \rho(r) = 
	\rho_{00}\left\{\mathrm{cosh}\left( \frac{r-R_\mathrm{c}}{H_0} \right)\right\}^{-2},
      \label{den prof}
\end{equation}
where $R_\mathrm{c}$ and $\rho_{00}$ are the radius and the 
density where $\mathrm{d} \rho /\mathrm{d} r=0$, respectively \citep[c.f.][]{S42}, and 
$H_0\equiv c_\mathrm{s}/\sqrt{2\pi G \rho_{00}}$ is the scale height.

From Equation (\ref{den prof}), 
if we determine $\rho_{00}$ and $R_\mathrm{c}$, the density profile
is completely specified except for the boundaries that are discussed later.
Here, the value of $R_\mathrm{c}$ itself loses
its physical meaning since the curvature is neglected. 
Therefore, only $\rho_{00}$ specifies the density profile.
The peak density $\rho_\mathrm{00}$ is determined by
the condition of the force balance at $r=R_\mathrm{CD}$.
The gravitational force must vanish at $r=R_\mathrm{CD}$  
because the total mass of the hot bubble is negligible.
Therefore, from Equation (\ref{hydrostatic}), the inner boundary conditions 
are given by
\begin{equation}
	\frac{c_\mathrm{s}^2}{\rho}\frac{\mathrm{d} \rho}{\mathrm{d} r}\Bigr|_{r=R_\mathrm{CD}} = g_\mathrm{dec}.
	\label{bound}
\end{equation}
The column density from $R_\mathrm{CD}$ to $R_\mathrm{c}$ is 
\begin{equation}
	\Sigma_\mathrm{dec}= \int_{R_\mathrm{CD}}^{R_\mathrm{c}} \rho \mathrm{d} r = 
	\rho_{00} H_0 \tanh \left( \frac{ R_\mathrm{c}-R_\mathrm{CD} }{H_0} \right)
	= \frac{g_\mathrm{dec}}{4\pi G},
	\label{sigmaacc}
\end{equation}
where we use Equation (\ref{bound}) in the last equality.
The characteristic column density $\Sigma_\mathrm{dec}$ represents the amount of 
the deceleration.
The ratio of 
the column density $\Sigma_\mathrm{s}$ to $\Sigma_\mathrm{dec}$ 
determines the importance of self-gravity relative to deceleration. 
From Equation (\ref{sigmaacc}) and the pressure equilibrium at the CD, 
$\rho(R_\mathrm{CD})c_\mathrm{s}^2 
=P_\mathrm{II}$, the peak density can be expressed by $\Sigma_\mathrm{dec}$ and $P_\mathrm{II}$ as follows:
\begin{equation}
	\rho_{00}c_\mathrm{s}^2 = P_\mathrm{II} + 2\pi G \Sigma_\mathrm{dec}^2.
	\label{den p sig}
\end{equation}
Substituting Equation (\ref{sigmaacc}) into Equation (\ref{den p sig}), 
one obtains
\begin{equation}
	\rho_{00}c_\mathrm{s}^2 = P_\mathrm{II} + \frac{g_\mathrm{dec}^2}{8\pi G}.
	\label{den00}
\end{equation}
The peak density $\rho_{00}$ is determined by the following way. 
We use the thin-shell model shown in Section \ref{sec:HII} to get 
$\mathrm{d}^2R_\mathrm{s}/\mathrm{d}t^2=-g_\mathrm{dec}$ and
${R}_\mathrm{s}$ at
any given times.
The pressure of the HII region $P_\mathrm{II}$ is given by Equation (\ref{Pii}).
Substituting obtained $g_\mathrm{dec}$ and $P_\mathrm{II}$ into Equation (\ref{den00}), 
we can get $\rho_{00}$, and can specify the functional form of the density profile.

Next, we determine the positions of boundaries, both the CD and the SF.
As mentioned above, only the distance relative to $R_\mathrm{c}$ has physical meaning.
The position of the CD and the SF is determined 
from the pressure balances on both sides which are given by 
\begin{equation}
 c_\mathrm{s}^2 \rho(R_\mathrm{CD}) = P_\mathrm{II}\;\;\;\mathrm{and}\;\;\;
 c_\mathrm{s}^2 \rho(R_\mathrm{SF}) = \rho_\mathrm{E} 
 \left( \frac{\mathrm{d} R_\mathrm{s}}{\mathrm{d}t} \right)^2,
        \label{bound0}
\end{equation}
respectively, where $\mathrm{d} R_\mathrm{s}/\mathrm{d}t$ is obtained from the thin-shell model.

\subsection{Three Evolutionary Phases of Density Profiles}\label{sec:den evo}
\begin{figure*}[htpb]
   \begin{center}
     \begin{tabular}{ccc}
         \includegraphics[width=5.5cm]{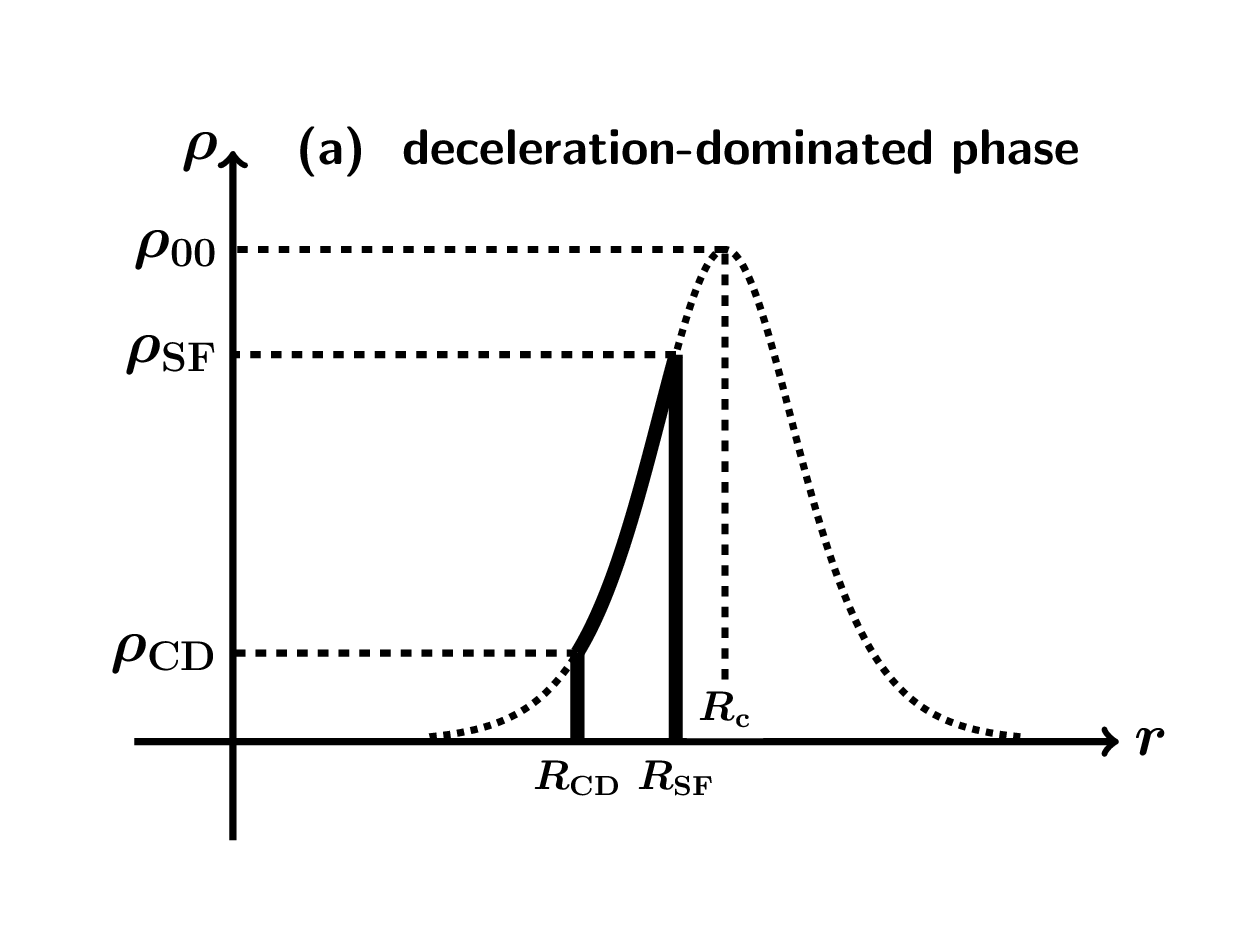}
         \hspace{-6mm}
         &
         \includegraphics[width=5.5cm]{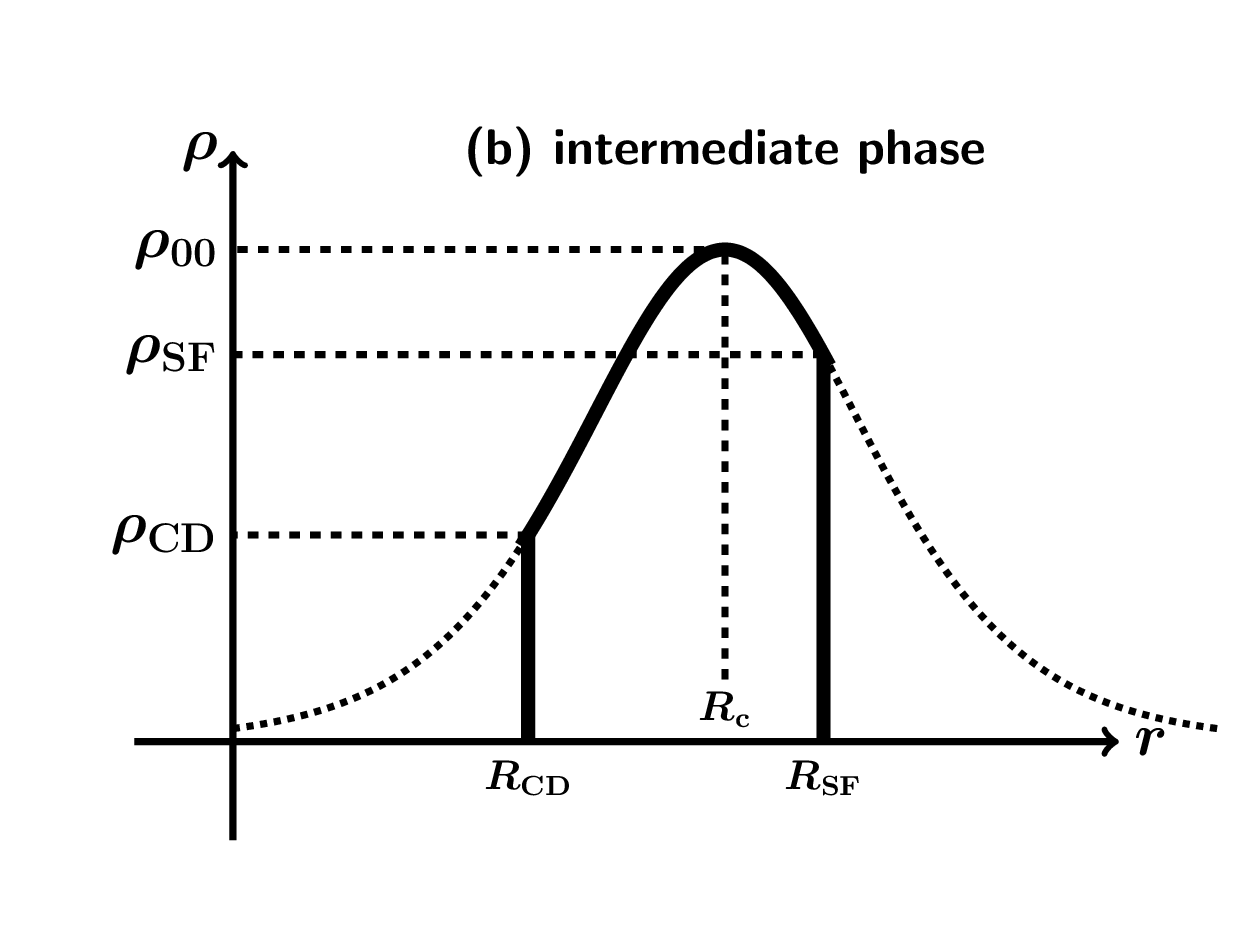}
         \hspace{-6mm}
         &
         \includegraphics[width=5.5cm]{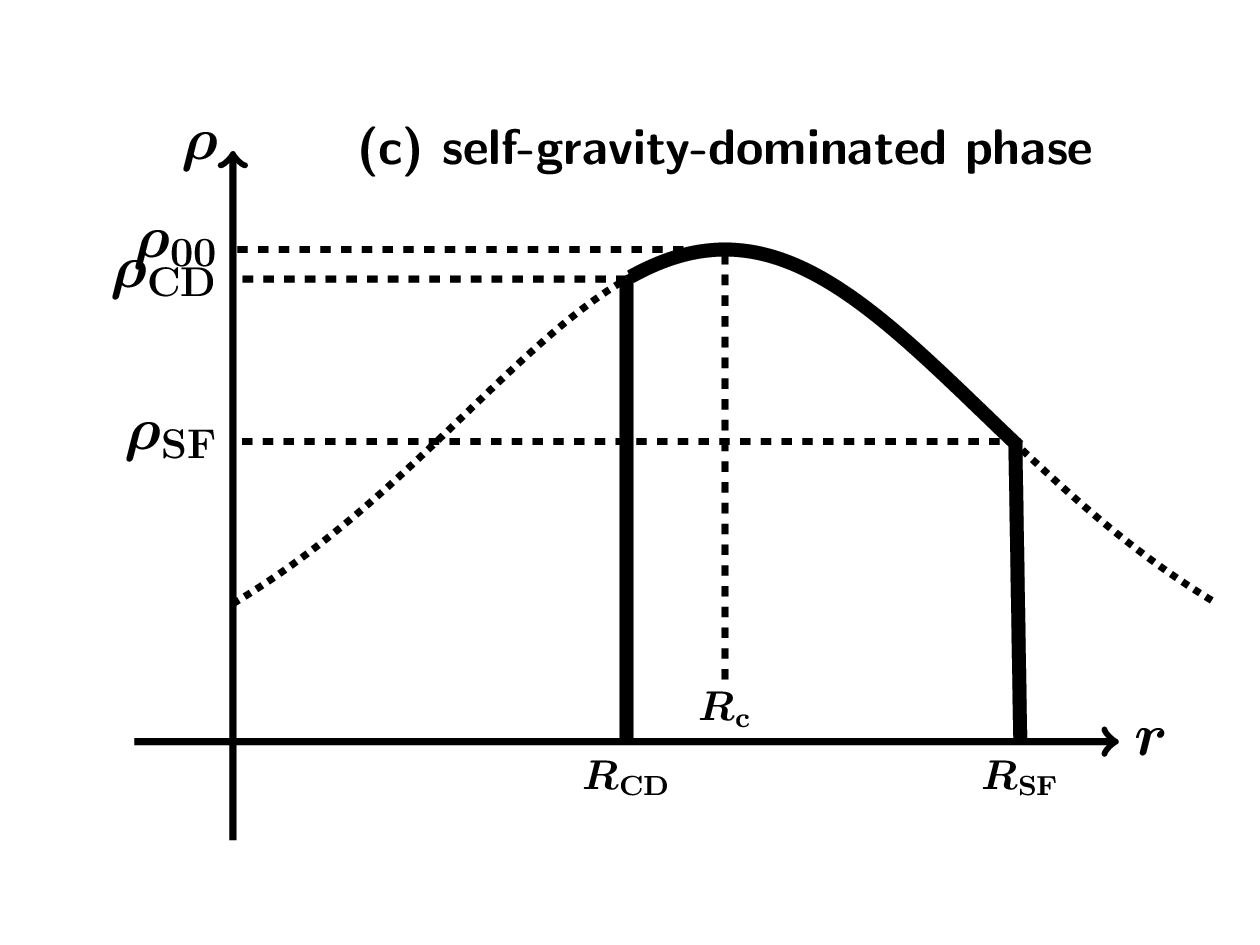}
     \end{tabular}
   \end{center}
\caption{
Schematic pictures of the density profiles of the shell
in (a) deceleration-dominated phase $(\Sigma_\mathrm{s}<\Sigma_\mathrm{dec})$,
(b) intermediate phase $(\Sigma_\mathrm{dec}<\Sigma_\mathrm{s}<2\Sigma_\mathrm{dec})$, 
and (c) self-gravity-dominated phase $(2\Sigma_\mathrm{dec}<\Sigma_\mathrm{s})$.
}
\label{fig:linear}
\end{figure*}
\begin{figure}[htpb]
        \begin{center}
             \includegraphics[width=8cm]{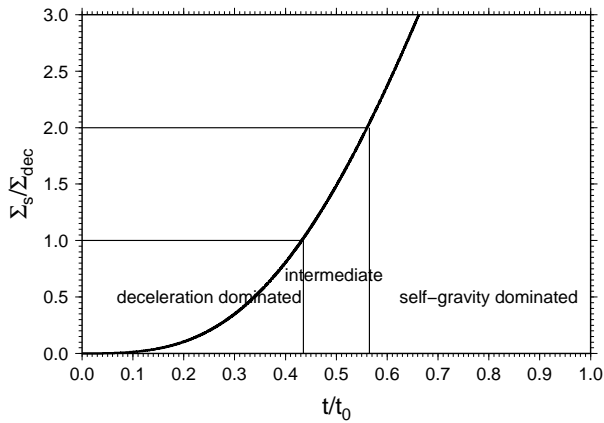}
        \end{center}
\caption{
Time evolution of the ratio of the column density $\Sigma_\mathrm{s}$ to the 
characteristic column density $\Sigma_\mathrm{dec}$ 
(Equation (\ref{sigmaacc})).
Evolutionary phases corresponding to Figure \ref{fig:linear} are labeled.
}
\label{fig:den_pro}
\end{figure}

The time evolution of the density profile is characterized by the ratio of 
the column density $\Sigma_\mathrm{s}$
to the characteristic column density $\Sigma_\mathrm{dec}$. 
Here, the column density $\Sigma_\mathrm{s}$ is given by 
the thin-shell approximation (see Section \ref{sec:HII}).
The column density $\Sigma_\mathrm{s}=\rho_\mathrm{E} R_\mathrm{s}/3$ can be derived from the 
mass conservation, since $M_\mathrm{s}=4\pi G R_\mathrm{s}^2\Sigma_\mathrm{s}$ (see Section \ref{sec:HII}).
The time evolution of the density profile 
is roughly divided into the following three phases: 
deceleration-dominated phase ($\Sigma_\mathrm{s}<\Sigma_\mathrm{dec}$),
intermediate phase ($\Sigma_\mathrm{dec}<\Sigma_\mathrm{s}<2\Sigma_\mathrm{dec}$), and 
self-gravity-dominated phase ($2\Sigma_\mathrm{dec}<\Sigma_\mathrm{s}$), depending on 
the value of $\Sigma_\mathrm{s}/\Sigma_\mathrm{dec}$. 
The schematic pictures of the density profiles in 
these three phases are shown in Figures \ref{fig:linear}.
Figure \ref{fig:den_pro} shows the time evolution of $\Sigma_\mathrm{s}/\Sigma_\mathrm{dec}$. 
In the early deceleration-dominated phase, 
$R_\mathrm{c}$ is outside of the shell and it is in front of the SF, 
or $R_\mathrm{SF}<R_\mathrm{c}$ (see Figure \ref{fig:linear}(a)).
This means that the actual density peak exists at $R_\mathrm{SF}$.
As the shell expands, $\Sigma_\mathrm{s}$ increases by accretion while $\Sigma_\mathrm{dec}$ decreases
by deceleration.
In Figure \ref{fig:den_pro}, it is seen that 
$\Sigma_\mathrm{s}$ becomes larger than $\Sigma_\mathrm{dec}$ at $t/t_0\sim 0.44$.
When $\Sigma_\mathrm{s}>\Sigma_\mathrm{dec}$, $R_\mathrm{c}$
is inside the shell.
In the intermediate phase ($\Sigma_\mathrm{dec}<\Sigma_\mathrm{s}<2\Sigma_\mathrm{dec}$), 
$R_\mathrm{c}$ is closer to $R_\mathrm{SF}$ than $R_\mathrm{CD}$ as shown in Figure \ref{fig:linear}(b). 
When $\Sigma_\mathrm{s}$ becomes larger than $2\Sigma_\mathrm{dec}$ ($t/t_0>0.57$), $R_\mathrm{c}$
becomes closer to $R_\mathrm{CD}$ than $R_\mathrm{SF}$ (see Figure \ref{fig:linear}(c)).
Since the period of the intermediate phase is relatively short,
roughly speaking,
the density profile transforms from the deceleration-
to the self-gravity-dominated profiles around $t/t_0\sim0.5$.

\subsection{Comparison with One-Dimensional Simulation}
Obtained density profile by above semi-analytic method 
is compared with results of 1D simulation.
We use the 1D spherically symmetric Lagrangian 
Godunov method \citep{vL97}.
We do not calculate the 
radiative transfer of ionizing photons and ionized gas, but the 
cold gas is pushed by interior pressure whose value is given by Equation (\ref{Pii}).
The equation of state is assumed to be isothermal.
We calculate the expanding shell around the 
41$M_{\odot}$ star 
that is embedded by the uniform ambient gas of $n_\mathrm{E}=10^3$ cm$^{-3}$.

\begin{figure}[htpb]
         \begin{center}
            \includegraphics[width=8cm]{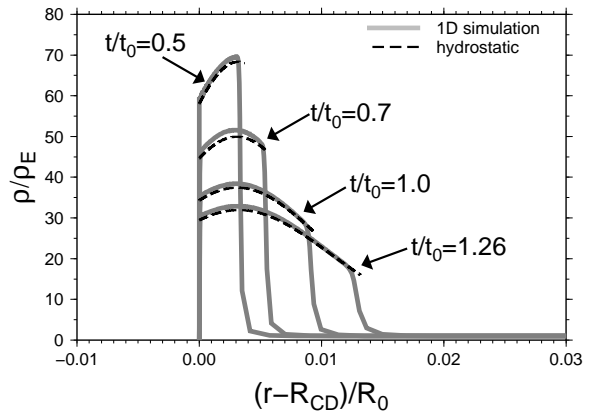}
         \end{center}
\caption{
Snapshots of density profiles for
$t/t_0=0.5$, 0.7, 1.0, and 1.26.
The abscissae are the distance from the CD.
The thick gray lines in the upper panel indicate the results of the 
1D simulation.
The dashed lines in the upper panel represent the 
instantaneous hydrostatic density profiles. 
}
\label{fig:1dim}
\end{figure}
Figure \ref{fig:1dim} shows the snapshots of density profiles for
$t/t_0=0.5$, 0.7, 1.0, and 1.26.
The thick gray lines represent the results of the 1D calculation.
The dashed lines 
show the density profiles obtained from the semi-analytic method.
It is seen that the semi-analytic method describes 
the density profile of the 1D simulation reasonably well. 
The density profile in the semi-analytic method is 
slightly lower than the results of the 1D calculation
because the actual CD expands a little slower
than the mean radius of the shell $R_\mathrm{s}$ in the thin-shell approximation.
As shown in Section \ref{sec:den evo}, one can see that 
the density peak moves the CD from the SF owing to the self-gravity.

\subsection{Scaling Law of Unperturbed Density Profiles}\label{sec:scaling den}
As shown in Section \ref{sec:HII}, 
the non-dimensional position of the shell, $\tilde{R}_\mathrm{s}$, is approximately independent
of the model parameters ($n_\mathrm{E}$, $Q_\mathrm{UV}$).
Similarly, it is useful to investigate how the 
density profile depends on the above parameters.
The non-dimensional pressures at the CD and the SF are given 
by Equation (\ref{nondim PII}) and $\tilde{V}_\mathrm{s}^2$,
respectively, that is, they are independent of the parameters.
Moreover, the pressure at the density peak 
$\tilde{P}_{00} = \tilde{\rho}_{00}\tilde{c}_\mathrm{s}^2$ is also 
independent of the parameters as seen in Equation (\ref{den00}).
Thus, noting that 
the non-dimensional sound speed $\tilde{c}_\mathrm{s}=c_\mathrm{s} t_0/R_0$ is 
proportional to 
the reciprocal of the typical Mach number ${\cal M}_0=4R_0/(7t_0c_\mathrm{s})$,
where the factor of $4/7$ arises from Equation (\ref{thin-shell}),
we have the scaling laws of $\tilde{H_0}$, $\tilde{\rho}_{00}$, and 
$\tilde{t}_\mathrm{ff}$ given by 
\begin{equation}
        \tilde{H}_0 \propto \tilde{c}^2_\mathrm{s} \tilde{P}_{00}^{-1/2} 
     \propto   {\cal M}_0^{-2},
        \label{H sou}
\end{equation}
\begin{equation}
        \tilde{\rho}_{00} \propto \tilde{c}_\mathrm{s}^{-2} \tilde{P}_{00} 
        \propto {\cal M}_0^{2},
\end{equation}
and 
\begin{equation}
        \tilde{t}_\mathrm{ff} \propto \tilde{\rho}_{00}^{-1/2} \propto 
        \tilde{c}_\mathrm{s} \tilde{P}_{00}^{-1/2} \propto {\cal M}_0^{-1},
        \label{t sou}
\end{equation}
respectively, where $t_\mathrm{ff}\equiv 1/\sqrt{2\pi G \rho_{00}}$ is 
the free fall timescale of the shell.
As a result, 
it is found that the density profiles
for various set of $(n_\mathrm{E},\;Q_\mathrm{UV})$ 
are characterized by a single parameter, that is the 
typical Mach number,
\begin{equation}
        {\cal M}_0
        = \frac{4}{7}\frac{R_0}{c_\mathrm{s} t_0}= 7\;
      Q_\mathrm{UV,49}^{1/7}\:
      T_\mathrm{c,10}^{-1/2}\:
      n_\mathrm{E,3}^{-1/14},
  \label{mach}
\end{equation}
where $T_\mathrm{c,10}=T_\mathrm{c}/10$ K.

\section{Influence of Asymmetric Density Profile on Gravitational Instability}\label{sec:asym}
As shown in Section \ref{sec:unp},
the expanding shell has the highly asymmetric density profile and it is expected to influence the GI.
In this section, we investigate influences of the asymmetric density profile on the 
dispersion relation of the GI.
What we discuss here is to extend the classical stability analysis
of the GI in the symmetric layer with respect to 
the mid-plane \citep{GL65,EE78,LP93} to
the GI in the asymmetric layer.
The linear analysis in the incompressible limit has been investigated by \citet{V88}.

We take $z$-axis parallel to the thickness of the layer, and take $x$-axis as the 
transverse direction.
The density is assumed to peak at $z=0$, and 
positions of boundaries are $z_1$ and $z_2$ ($z_1<z_2$).
We consider a layer that is subject to a constant deceleration. 
The deceleration arises from the difference of pressures on the boundaries ($z=z_1$ and $z_2$).
In this case, the position of the density peak is not in the mid-plane of the layer, the 
density profile is asymmetric, or $-z_1 \ne z_2 $.
The amount of deceleration directly enhances the degree of asymmetry of the density profile. 

\begin{figure*}[htpb]
   \begin{center}
         \includegraphics[width=14.0cm]{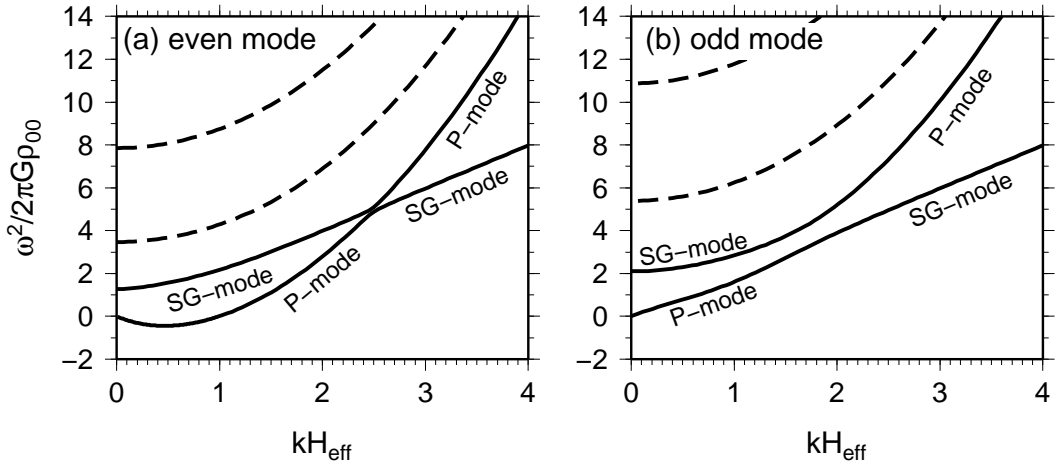}
   \end{center}
\caption{
Dispersion relations for (a) the even mode and (b) the odd mode in 
the symmetric layer with $|z_1|=z_2=3H_0$.
The ordinate denotes $\omega^2/2\pi G \rho_{00}$. 
The abscissa denotes the wavenumber multiplied by the effective thickness of the shell, $H_\mathrm{eff}
\equiv \sigma/(2\rho_{00})$.
The dashed lines show higher harmonics of the sound wave with respect to the $z$-direction.
}
\label{fig:sym3.0}
\end{figure*}
\begin{figure*}[htpb]
   \begin{center}
         \includegraphics[width=14.0cm]{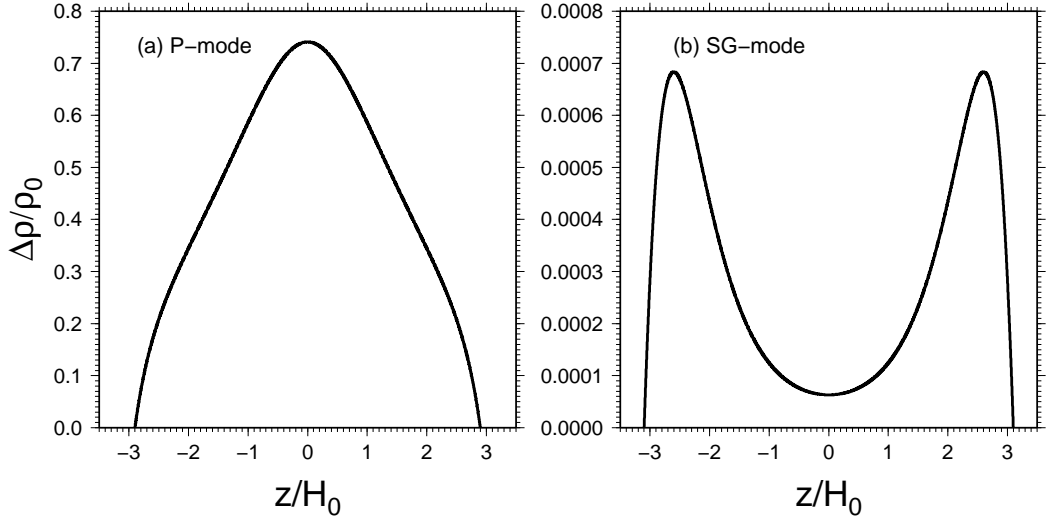}
   \end{center}
\caption{
Distribution of the Lagrangian density perturbation,
$\Delta \rho/\rho_0\equiv \delta \rho/\rho_0 + v_z \mathrm{d}\ln\rho_0(z)\mathrm{d}z$ for 
$kH_\mathrm{eff}=3.5$ in the even mode. Each panel corresponds to 
(a) the P-mode and (b) the SG-mode.
The normalization is determined by $|\delta z_2|/z_2=0.1$.
}
\label{fig:sym3.0 per}
\end{figure*}

\subsection{Perturbation Equations}
We consider the following perturbations:
\begin{eqnarray}
  \rho(z,x,t) & = & \rho_0(z) + \delta \rho(z)e^{i(kx - \omega t)}, \nonumber \\
  v_z(z,x,t) & = & v_z(z)e^{i(kx - \omega t)},  \\
  v_x(z,x,t) & = & v_x(z)e^{i(kx - \omega t)},  \nonumber\\
  \phi(z,x,t) & = & \phi_0(z) + \delta \phi(z)e^{i(kx - \omega t)}. \nonumber
\end{eqnarray}
Perturbation equations are
\begin{equation}
	-i\omega\delta \rho + \frac{\mathrm{d} (\rho_0 v_z)}{\mathrm{d} z} +
      \rho_0 ik v_x = 0,
      \label{eoc per}
\end{equation}
\begin{equation}
	i\omega v_z = \frac{\mathrm{d}}{\mathrm{d} z}
	\left( c_\mathrm{s}^2\frac{\delta \rho}{\rho_0} + \delta \phi \right),
      \label{eomr per}
\end{equation}
\begin{equation}
        \omega v_x = k
	\left( c_\mathrm{s}^2\frac{\delta \rho}{\rho_0} + \delta \phi \right),
      \label{eomy per}
\end{equation}
and
\begin{equation}
        \frac{\mathrm{d}^2 \delta \phi }{\mathrm{d} z^2} - k^2 \delta \phi = 
	4\pi G \delta \rho,
      \label{poi per}
\end{equation}
where the sound speed is assumed to be constant.

\subsubsection{Boundary Conditions}
To concentrate on the effect of asymmetry of the unperturbed state, we impose the 
CD boundary conditions 
at both $z_1$ and $z_2$ \citep{GL65,EE78} as follows:
\begin{equation}
        \delta \rho(z_1) = - \frac{\mathrm{d} \rho_0}{\mathrm{d} z}\Bigr|_{z=z_{1}}\delta z_1,\;\;
    v_z(z_1) = -i\omega \delta z_1,
    \label{bou 1 z1}
\end{equation}
\begin{equation}
 \frac{\mathrm{d} \delta \phi}{\mathrm{d} z}\Bigr|_{z=z_{1}} - k\delta \phi(z_1) + 4\pi G \rho_0(z_1) \delta z_1,
    \label{bou 2 z1}
\end{equation}
and 
\begin{equation}
        \delta \rho(z_2) = - \frac{\mathrm{d} \rho_0}{\mathrm{d} z}\Bigr|_{z=z_{2}}\delta z_2,\;\;
    v_z(z_2) = -i\omega \delta z_1,
    \label{bou 1 z2}
\end{equation}
\begin{equation}
 \frac{\mathrm{d} \delta \phi}{\mathrm{d} z}\Bigr|_{z=z_{2}} + k\delta \phi(z_2) + 4\pi G \rho_0(z_2) \delta z_2,
    \label{bou 2 z2}
\end{equation}
where $\delta z_1$ and $\delta z_2$ are the displacements of the surfaces at $z_1$ and $z_2$, respectively.

\subsubsection{Numerical Methods}\label{sec:numerical}
We solve Equations (\ref{eoc per})-(\ref{poi per}) as a boundary-value problem for a given 
wavenumber.  Equations (\ref{eoc per}), (\ref{eomr per}), and (\ref{poi per}) are 
integrated from $z=z_1$ to $z_2$
by using the fourth order Runge-Kutta method.
Note that $v_x$ is determined by $\delta \rho$ and $\delta \phi$ from Equation (\ref{eomy per}). 
Given $\omega$, at $z=z_1$, we have five unknown variables ($\delta \rho$, $v_z$, $\delta \phi$, 
$\mathrm{d}\delta\phi/\mathrm{d}z$, and $\delta z_1$), and 
have three boundary conditions (see Equations (\ref{bou 1 z1}) and (\ref{bou 2 z1})).
Therefore, if we determine two variables $Q_1$ and $Q_2$, all variables at $z_1$ are specified, 
where $Q_1$ is one of ($\delta \rho$, $v_z$, $\delta z_1$) and $Q_2$ is one of ($\delta \phi$, 
$\mathrm{d}\delta\phi/\mathrm{d}z$).
Generally, the boundary conditions at $z_2$ are not satisfied if we start from arbitrary values 
of $Q_1$ and $Q_2$ at $z_1$.
Equation (\ref{bou 1 z2}) can always be satisfied by using a linear combination of 
two independent solutions having the boundary values $(Q_1(z_1),\;Q_2(z_1))=(1,0)$ and (0,1). 
Eigenvalue, $\omega$, is modified iteratively until the solutions satisfy Equation 
(\ref{bou 2 z2}) by using the Newton-Raphson method.

\subsection{Symmetric Layer}
First, we investigate the symmetric case with $-z_1=z_2=3H_0$. Because of symmetry,
perturbation can be divided by even and odd modes completely. 
In the even (odd) mode, the density perturbation is symmetric (antisymmetric) with respect to the mid-plane.
Figures \ref{fig:sym3.0}(a) and (b) show the dispersion relations for the even and odd modes, 
respectively.
The abscissa denotes the wavenumber multiplied by the effective thickness of the shell, $H_\mathrm{eff}
\equiv \sigma/(2\rho_{00})$.
It is well known that the unstable mode ($\omega^2<0$) is found only in the even mode.
The unstable mode belongs to the ``compressible mode'' that means that 
the density perturbation in the central region collapses leaving behind the gas around boundaries.
The detailed structure of stable modes is also plotted in Figure \ref{fig:sym3.0}.
The stable mode can be divided by the ``P mode'' (pressure mode, or compressible mode) and 
the ``SG mode'' (surface-gravity mode).
Figure \ref{fig:sym3.0 per} shows that the distribution of the Lagrangian density perturbation $\Delta \rho\equiv\delta\rho 
+ v_z \mathrm{d}\rho_0/\mathrm{d}z$ for $kH_\mathrm{eff}=3.5$ in the even mode.
Figures \ref{fig:sym3.0 per}(a) and (b) correspond to the P and the SG modes, respectively.
The normalization is determined by $|\delta z_2|/z_2=0.1$.
One can see that $\Delta \rho$ profiles of the P and SG modes are quite different.
In the P mode, $\Delta \rho/\rho_0$ peaks at the mid-plane. 
The displacement of the boundary $|\delta z_2|/z_2$ is negligible
compared with $\Delta \rho/\rho_0$. The P mode propagates as longitudinal variation of pressure.
On the other hand, the SG-mode has two $\Delta \rho/\rho_0$ peaks near both boundaries, and 
has the minimum value at the mid-plane.
Moreover, since $|\delta z_2|/z_2$ is much larger than $\delta \rho/\rho_0$,
the SG mode is almost incompressible.
The SG mode propagates as the deformation of the surface.
In Figure \ref{fig:sym3.0}(a), it is seen that the unstable mode transforms into the stable 
P mode around $kH_\mathrm{eff}\sim1$.
On the other hand, there is the another stable mode labelled by the SG mode ($kH_0<2$).
The P and SG modes approach each other as the wavenumber rises from the small limit. 
One can see a remarkable feature around $kH_\mathrm{eff}\sim2.5$ where
these two modes do not intersect but begin to move apart.
At this point, they exchange their properties, 
suggesting the mode exchange.
The mode exchange between the P and the SG modes 
is also occurred in the odd mode (see Figure \ref{fig:sym3.0}(b)).
This is the first time when the mode exchange is found in the dispersion relation of 
self-gravitating layer.
The dashed lines in Figures \ref{fig:sym3.0} show higher harmonics of the 
sound wave with respect to the $z$-direction.
For large wavenumber, the frequencies of the P and the 
SG modes show different dependence on wavenumber.
The P mode shows $k^2$ dependence while the SG mode shows $k$ dependence.
The angular frequencies of the SG modes associated by the deformation of $z_1$ and $z_2$ are
\begin{equation}
        \omega_\mathrm{SG,z_1}^2 = - 2\pi G \rho_0(z_1) 
        + \frac{1}{\rho_0}\frac{\partial P_0}{\partial r} \Bigr|_{z_1}k
        \label{sg mode}
\end{equation}
and 
\begin{equation}
        \omega_\mathrm{SG,z_2}^2 = - 2\pi G \rho_0(z_2) 
        - \frac{1}{\rho_0}\frac{\partial P_0}{\partial r} \Bigr|_{z_2}k,
\end{equation}
respectively \citep{WS81}, for $kH_0\gg1$.
The SG branches for large wavenumber in the even and odd modes are identical with each other because the 
surface gravities at the two boundaries are the same.
\begin{figure*}[htpb]
        \begin{center}
              \includegraphics[width=14.0cm]{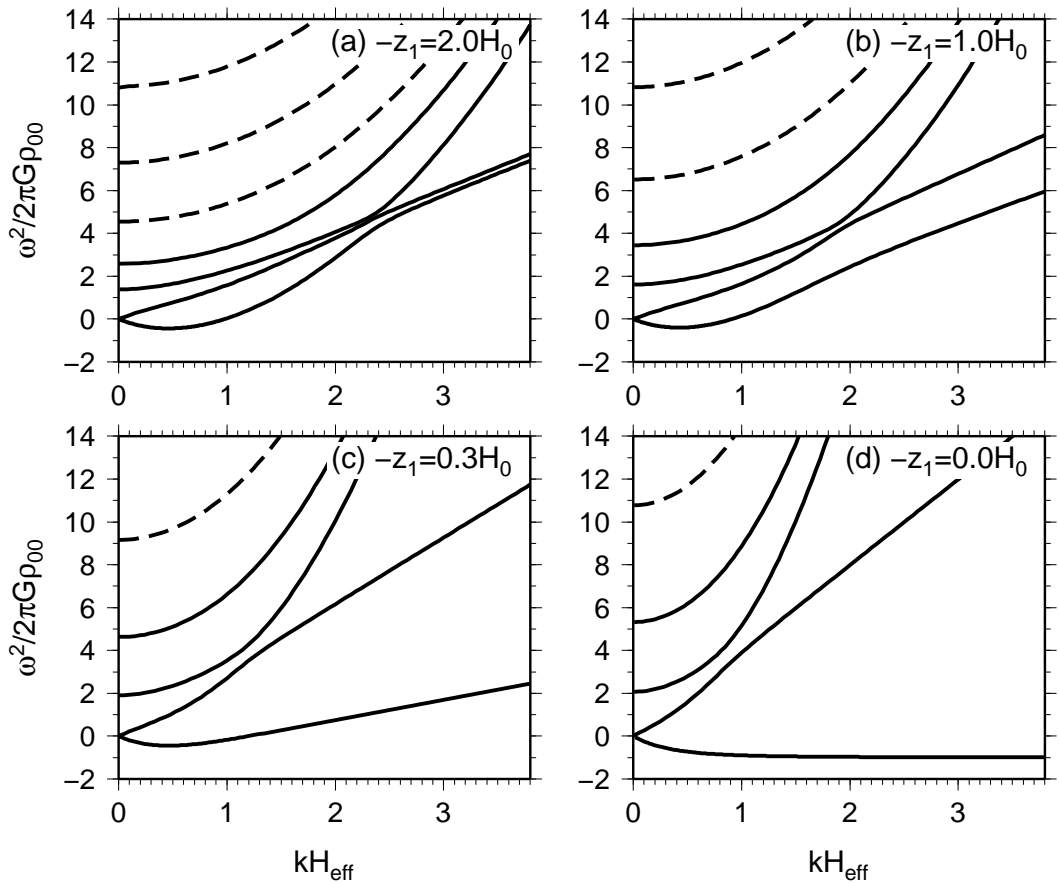}
        \end{center}
\caption{
Dispersion relations for $-z_1/H_0=$(a)2.0, (b)1.0, (c)0.3, and (d)0.0.
The ordinate and the abscissa are the same as Figure \ref{fig:sym3.0}.
}
\label{fig:asym}
\end{figure*}
\begin{figure}
   \begin{center}
         \includegraphics[width=8.0cm]{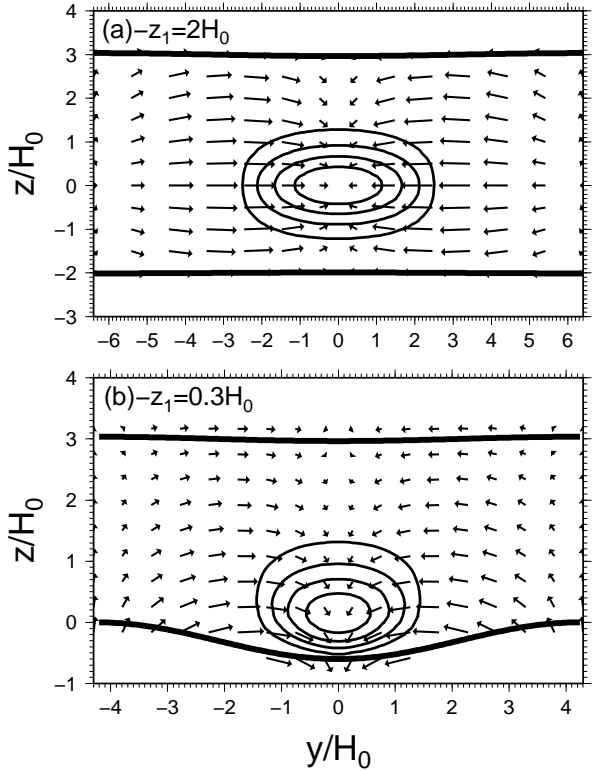}
   \end{center}
\caption{
Cross sections of the layers with the fastest growing mode 
for (a)$-z_1=2.0H_0$ and (b)$-z_1=0.3H_0$.
The contour indicates the density perturbation normalized by $\rho_{00}$.
The contour levels of the density perturbation take values of 0.04, 0.08, and 0.16. 
}
\label{fig:asym eigen}
\end{figure}
\subsection{Asymmetric Layer}
In this section, we investigate the dependence of the dispersion relation 
on the degree of the asymmetry by changing $z_1$.
Since the layer is no longer symmetric with respect to $z=0$, 
perturbations cannot be divided into 
the even and the odd modes.
Figure \ref{fig:asym}(a) shows the dispersion relation for $z_1=-2H_0$.
One can see more complex structure of the mode exchanges around $kH_\mathrm{eff}\sim2.4$
than that in Figure \ref{fig:sym3.0}.
For large wavenumber, the angular frequencies of the two SG modes split because 
$\omega_\mathrm{SG,z_1}^2<\omega_\mathrm{SG,z_2}^2$.
Figure \ref{fig:asym eigen}(a) shows the cross section of the layer ($z_1=-2H_0$) in
the fastest growing mode. The contour indicates the density perturbation 
normalized by $\rho_{00}$.
Here, we take $\delta \rho_\mathrm{max}/\rho_{00}=0.2$ to specify the 
normalization of the perturbations.
The arrows represent the velocity vectors.
The boundary surfaces hardly deform, and 
the gas collapses from all directions to the center ($z=0$, $x=0$).
This behavior corresponds to the compressible mode.
In Figure \ref{fig:asym}(a), 
the unstable branch transforms the P mode around $kH_\mathrm{eff}\sim1$ and 
it is connected with the SG mode around $kH_\mathrm{eff}\sim2.4$
through the mode exchange. 
The case with stronger asymmetry with $z_1=-H_0$ is shown in Figure \ref{fig:asym}(b).
In this case, the difference of the angular frequencies between the two SG modes is larger 
because the surface gravity at $z_1$ is lower.
The frequency of the SG mode associated with $z_1$ becomes smaller than that with $z_2$.
Comparing with Figure \ref{fig:asym}(a), as well as frequency, 
the wavenumber of the mode exchange is smaller. 
As a result, the frequency range of P mode is narrower and the range of the SG mode spreads.
The P mode $\omega^2 \propto k^2$ is expected to disappear when the wavenumber of the 
mode exchange is smaller than a critical wavenumber that separates unstable mode from stable mode.

The case with $-z_1<H_0$ is quite different from the case with $-z_1\ge H_0$.
The dispersion relation for $-z_1=0.3H_0$ is shown in Figure \ref{fig:asym}(c). 
In Figure \ref{fig:asym}(c), one can see that
the angular frequency of the SG mode $\omega_\mathrm{SG,z_1}^2$ 
is significantly lower than $\omega_\mathrm{SG,z_1}^2$ for large wavenumbers. 
Unlike the case with $-z_1>H_0$, 
the unstable mode appears to directly connect with the 
SG mode around $kH_\mathrm{eff}\sim1.2$ as mentioned above. 
Figure \ref{fig:asym eigen}(b) shows the cross section of the layer ($z_1=-0.3H_0$) 
in the fastest growing mode.
The gas tends to collect toward the density peak $z=0$ because the unperturbed 
gravitational potential has the minimum value there.
One can see that eigen-functions in $z>0$ are similar to those 
in Figure \ref{fig:asym eigen}(a).
The gas collapses toward the center ($z=0$, $x=0$) leaving behind the gas around $z_2$.
However, in the region where $z<0$, eigenfunctions are quite different.
The sound wave can travel between $z_1$ and the density peak many times 
during the development of the GI.
Therefore, collapse toward $z=0$ is suppressed by the pressure gradient.
However, the GI can proceed even in $z<0$ through 
the deformation of the $z_1$ that
makes the gravitational potential deeper.
From Figure \ref{fig:asym eigen}(b), 
one can see that the velocity field is not headed for the density peak ($z=0)$ but arises 
so that the surface at $z_1$ deforms. 
Therefore, the features of GI in the region $z>0$ and $z<0$ 
have properties of ``compressible mode'' and 
``incompressible mode'', respectively.
If the distance of the $z_1$ from the density peak is zero, 
the layer is unstable for all wavenumbers (see Figure \ref{fig:asym}(d)).

\begin{figure}[htpb]
        \begin{center}
         \includegraphics[width=8.0cm]{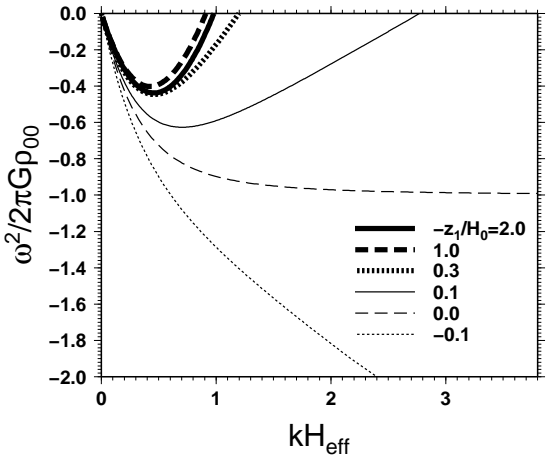}
        \end{center}
\caption{
Dispersion relation of the asymmetric layer for 
$-z_1/H_0=2$(the thick solid line), $1$(the thick dashed line), 
$0.3$(the thick dotted line), $0.1$(the thin solid line), 
$0$(the thin dashed line), and $-0.1$(the thin dotted line), 
where $z_2/H_0$ is assumed to be $3.0$.
}
\label{fig:unstable}
\end{figure}

Figure \ref{fig:unstable} shows the dispersion relation of the unstable mode for 
variety of $z_1$ with $z_2=3H_0$.
The thick solid and the thick dashed lines correspond to $-z_1/H_0=2$ and 1, 
respectively.  For $-z_1/H_0\ge1$, the growth rate $-\omega^2/2\pi G\rho_{00}$ 
decreases as $-z_1/H_0$ decreases.
This property is the same as that in symmetric layers \citep[e.g., see Figure 1 in][]{NIM98}.
On the other hand, for the cases with $-z_1/H_0=0.3$ and 0.1, 
Figure \ref{fig:unstable} shows that 
the maximum growth rate increases as $-z_1/H_0$ decreases, indicating 
the opposite tendency to the case with $-z_1/H_0>1$.
On the other hand, the wavenumber of the most unstable mode is not different so much.
One can see that the square growth rate in large wavenumber is proportional to $\propto k$ while 
the square growth rate for $-z_1/H_0>1$ is proportional to $\propto k^2$.
The unstable mode appears to directly connect with the surface gravity mode at $z_1$ whose 
frequency is given by Equation (\ref{sg mode}). 
This is also seen in Figure \ref{fig:disp} for $k>k_\mathrm{max}$.
For $-z_1/H_0=0$, the surface gravity mode at $z_1$ becomes unstable for all wavenumber.
In this case, destabilized surface gravity wave  has
the growth rate 
$\omega_\mathrm{SG,z_1}^2 = -2\pi G \rho(z_1)<0$ independent of $k$ for large wavenumber limit
(see Equation (\ref{sg mode})).
This is the case with a static shell with $g_\mathrm{dec}=0$ (Equation (\ref{hydrostatic})).
\citet{TI83} investigated this situation including shell curvature and found 
that the shell is unstable for all wavenumber \citep[also see][]{WS81}.
For $-z_1/H_0=-0.1$, 
the square growth rate increases as $\propto k$ with wavenumbers because 
$\omega_\mathrm{SG,z_1}^2 \sim \frac{1}{\rho_0}\frac{\partial P_0}{\partial r} 
\Bigr|_{z_2}k<0$ (Equation (\ref{hydrostatic})).
This is well-known scaling law of the Rayleigh-Taylor instability.
The enhancement of the growth rate for $-z_1/H_0<1$ arises from the combination of the GI 
and the Rayleigh-Taylor instability.

\section{Gravitational Instability of Expanding Shells}\label{sec:shell}
In previous section, we focus on the effect of asymmetry of the 
density profile by imposing 
the same boundary conditions in both boundaries.
In this section, in a more realistic situation, 
we investigate the stability of expanding shells driven by the expansion the HII region.
The unperturbed density profile at each instant of time 
is given by the semi-analytic method presented in Section \ref{sec:unp}.
We neglect the curvature effect and
solve the perturbation Equations (\ref{eoc per})-(\ref{poi per}) but $z\rightarrow r$.
In this case, as well as in the asymmetric density profile, 
the difference of boundary properties between leading (the SF) and trailing (the CD) 
surfaces plays important roles in the GI.

\subsection{Influences of Boundaries on the Gravitational Instability}
\label{sec:influence}
Before presenting a linear analysis, we review how the 
SF and the CD influence the GI through the boundary effect. 
This point is important in understanding the results of the linear analysis.
In the early phase, the shell is highly confined by the ram pressure 
on the leading surface and 
by the thermal pressure on the trailing surface.
In this phase, 
the pressure at boundaries is as large as that at the density maximum, and 
the thickness of the shell is much smaller than 
the scale height, $H_0=c_\mathrm{s}/\sqrt{2\pi G \rho_{00}}$, where $\rho_{00}$ is the 
maximum density. Thus, in this phase, the boundary effect can strongly influence the GI. 
In the later phase, the boundary effect of the CD 
is expected to be also important because the density peak is close to the CD
as shown in Section \ref{sec:den evo}.
In this section, we summarize how the growth rate of GI is controlled by the 
different boundary conditions.
For simplicity, in Section \ref{sec:influence}, the layer is assumed to be symmetric 
with respect to the mid-plane, and 
physical variables are averaged across the thickness.

\subsubsection{Shock-confined Layer}\label{sec:shock}
Many authors have investigated influences of the SF on the GI 
\citep{V83,E89,N92,V94,E94,Wetal94a,IT08}.
The dispersion relation of the shock-confined layer is given by 
\begin{equation}
        \omega^2 \simeq c_\mathrm{s}^2k^2 -  2\pi G k\Sigma_\mathrm{s},
 \label{disp shock}
\end{equation}
where $\Sigma_\mathrm{s}$ is the column density, and  
$k$ is the transverse wavenumber of the perturbation.
This dispersion relation is the same as that for 
the infinitesimally thin layer.
In the highly confined layer, it is well known that 
the perturbation behaves like incompressible mode 
because the sound-crossing time over the thickness is much smaller 
than the free-fall timescale, $\sim1/\sqrt{G\rho_{00}}$ \citep{EE78,LP93}.
Therefore, density fluctuation is small.
The layer becomes unstable mainly by the deformation of the surfaces that
makes the perturbation of the gravitational potential deeper.
Hereafter, we call this mode the ``incompressible mode''.
The deformation of the SF generates the tangential flow
carrying the gas from the convex to the concave regions (seen 
from the downstream). 
Therefore, the tangential flow tends to make the SF flat, suggesting that 
it suppresses the growth of the GI.
In the shock-confined layer, the restoring term $c_\mathrm{s}^2 k^2$ arises
from the tangential flow behind the oblique SF.
On the other hand, in the case of the infinitesimally thin layer, 
this term comes 
from the pressure gradient.
Therefore, the origin of the restoring force is quite different.
From Equation (\ref{disp shock}), the maximum growth rate is given 
by $\pi G \Sigma_\mathrm{s}/c_\mathrm{s}$, and 
the corresponding wavenumber is given by $\pi G \Sigma_\mathrm{s}/c_\mathrm{s}^2$.
When $\Sigma_\mathrm{s}$ is small ($\le \rho_{00}H_0)$, the maximum growth rate is smaller than 
the inverse of the free-fall timescale, $\sqrt{G \rho_{00}}$, and 
the corresponding scale is larger than the scale height $H_0$ that is comparable to the Jeans scale.

\subsubsection{Pressure-confined Layer}
Next, we review the influence of the CD on the GI.
We consider the layer confined by thermal 
pressure of hot rarefied gases (CD boundary condition) on both sides. 
The dispersion relation becomes
\begin{equation}
        \omega^2 \simeq 2\pi G \Sigma_\mathrm{s} L_\mathrm{s} k^2 - 2\pi G k\Sigma_\mathrm{s},
 \label{disp pre}
\end{equation}
where $L_\mathrm{s}$ is the thickness of the layer, and  
we consider the large-scale limit where $k\ll 1/L_\mathrm{s}$.
Detailed derivation of Equation (\ref{disp pre}) 
is shown in Appendix 3 of \citet{IT08}.
In the pressure-confined layer, the stabilization effect of the tangential 
flow does not exist.
Therefore, the restoring term in Equation (\ref{disp pre}) is 
quite different from that in Equation (\ref{disp shock}).
Using the gravitational acceleration at the surfaces as
$|g|=2\pi G \Sigma_\mathrm{s}$, 
we can express the restoring term in Equation (\ref{disp pre}) 
by $|g|L_\mathrm{s} k^2$. 
Therefore, one can see that the restoring force arises from the 
surface gravity wave.
The layer with the CD boundary condition is less stabilized 
compared with the shock boundary condition because
the phase velocity of the gravity wave $\sqrt{|g|L_\mathrm{s}}$ is much smaller than 
$c_\mathrm{s}$ when $L_\mathrm{s}\ll H_0$. 
From the dispersion relation (\ref{disp pre}), 
the maximum growth rate is comparable to the inverse of the free fall 
time of the layer $\simeq \sqrt{G\rho_{00}}$, 
and the corresponding wavelength 
is about the thickness of the layer, $\simeq L_\mathrm{s}$ \citep{EE78,LP93}.
Therefore, one can see that the most unstable mode in the pressure-confined layer 
has a larger growth rate and a smaller scale than the shock-confined layer
\citep[see Figure 9 of][in detail]{IT08}.

\subsubsection{Expanding Shells}
Equations (\ref{disp shock}) and (\ref{disp pre}) cannot be applied directly in 
the GI of the expanding shells because the GI is expected to be stabilized 
by evolutionary effects, such as the expansion of the shell and 
the accretion of fresh gas through the SF.
\citet{E94} derived the following approximate dispersion relation,
\begin{equation}
i\omega = -\frac{3V_\mathrm{s}}{R_\mathrm{s}} 
    + \sqrt{\left( \frac{V_\mathrm{s}}{R_\mathrm{s}} \right)^2 
        + 2\pi G k\Sigma_\mathrm{s} - c_\mathrm{s}^2k^2}.
        \label{disp elme}
\end{equation}
The terms with $V_\mathrm{s}/R_\mathrm{s}$ come from evolutionary effects that 
stabilize the GI.

One can see that Equation (\ref{disp elme}) for the limit of $V_\mathrm{s}/R_\mathrm{s}\rightarrow0$ 
is the same as Equation (\ref{disp shock}).
Therefore, \citet{E94} and \citet{Wetal94b} essentially applied Equation (\ref{disp shock}) 
in the context of the GI of the expanding shell. 
However, they did not take into account the boundary effect of the CD on the trailing surface.
Comparing Equations (\ref{disp shock}) and (\ref{disp pre}), we suggest
that the stability of the thin shell neglecting 
the effect of the CD is suffered by large stabilizing effect, 
and it will underestimate the growth rate of GI in expanding shells.

\subsection{Boundary Condition}\label{sec:linear bc}
First, we assume that a constant pressure exerts on the CD all the time
 \citep[the CD boundary condition;][]{GL65,EE78}.
The boundary conditions are
\begin{equation}
        \delta \rho(R_\mathrm{CD}) 
        = - \frac{\mathrm{d} \rho}{\mathrm{d} r}\Bigr|_{r=R_\mathrm{CD}}\delta R_\mathrm{CD},\;\;\;
    v_r(R_\mathrm{CD}) = i\omega \delta R_\mathrm{CD},
	\label{boundary P}
\end{equation}
and 
\begin{equation}
        \frac{\mathrm{d} \delta \phi}{\mathrm{d} r} - k\delta \phi + 4\pi G \rho(R_\mathrm{CD}) \delta R_\mathrm{CD}=0,
	\label{boundary phi}
\end{equation}
where $\delta R_\mathrm{CD}$ is the displacement of the CD.

Next, let us consider the boundary conditions at $r=R_\mathrm{SF}$.
Since the unperturbed state is assumed to be the hydrostatic configuration, 
it is impossible to impose the shock boundary conditions self-consistently.
In order to treat it self-consistently, time-dependent
initial value problem is needed to be solved \citep{W82,IT08}.
Therefore, in this paper, we mimic the shock boundary conditions by 
introducing the stabilization effect.
We consider the following two approximate boundary conditions.

{\it Rigid surface boundary condition (RSBC). 
$\frac{\;\;\;\;\;\;}{\;\;\;\;\;\;}$}
\citet{V88} and \citet{UHF95} assumed that no ripples arise on the surface, or $\delta R_\mathrm{SF}=0$,
where $\delta R_\mathrm{SF}$ is the displacement of the SF.
The reason why we adopt $\delta R_\mathrm{SF}=0$ is that
the thin-shell linear analysis of the layer confined by rigid surfaces gives
the same dispersion relation as that of the shock-confined layer 
(Equation (\ref{disp shock})). 
In more precisely,
in the shock-confined layer, 
the tangential flow boosts the suppression effect against the self-gravity as 
mentioned in Section \ref{sec:shock}.
Instead, the RSBC weakens the self-gravity. 

{\it Tangential flow boundary condition (TSBC). $\frac{\;\;\;\;\;}{\;\;\;\;\;}$}
If the SF is rippled, the tangential flow behind the SF is generated.
Therefore, we set the tangential velocity $v_x$ at $r=R_\mathrm{SF}$.
Linearizing the Rankine-Hugoniot relation, we have
\begin{equation}
        v_x (R_\mathrm{SF}) = - \left( \dot{R}_\mathrm{SF} - \frac{c_\mathrm{s}^2}{\dot{R}_\mathrm{SF}} \right) 
        ik \delta R_\mathrm{SF}.
        \label{bound tan}
\end{equation}
The detailed derivation of Equation (\ref{bound tan}) is found in \citet{IT08}.

With both of above boundary conditions (RSBC and TFBC), we also impose 
the following ordinary used boundary conditions, 
\begin{equation}
    v_r(R_\mathrm{SF}) = i\omega \delta R_\mathrm{SF},
\end{equation}
and
\begin{equation}
        \frac{\mathrm{d} \delta \phi}{\mathrm{d} r} + k\delta \phi + 4\pi G \rho(R_\mathrm{SF}) \delta R_\mathrm{SF}=0.
\end{equation}

It is well known that the SF of the deceleration shell is subject to 
the hydrodynamical overstability \citep{V83}. 
The linear analysis in this paper cannot capture the Vishniac 
instability (VI) correctly since the 
approximate shock boundary conditions are imposed.
The effect of the VI is discussed in Section \ref{sec:discuss}.

The numerical method is the same as that in Section \ref{sec:numerical}.

\subsection{Scaling Law of Dispersion Relations}\label{sec:scaling disp}
As shown in Section \ref{sec:scaling den}, it is found that 
the density profiles are characterized by a single parameter ${\cal M}_0$.
This is because the scale height, the peak density, and the free fall time 
have the scaling laws with respect to 
${\cal M}_0$ as shown in Equations (\ref{H sou})-(\ref{t sou}).
The same is the case with the perturbation equations and the dispersion relation.
The non-dimensional maximum 
growth rate $\tilde{\omega}_\mathrm{max}\equiv\omega_\mathrm{max} t_0$ 
and the corresponding 
wavenumber $\tilde{k}_\mathrm{max}\equiv k_\mathrm{max}R_0$ scale as $\propto {\cal M}_0$
and $\propto {\cal M}_0^{2}$, respectively.
Therefore, in the present model, the evolution of the shell for various set 
of $(n_\mathrm{E},\:Q_\mathrm{UV})$ can be described by a single unperturbed
profile and a single time-dependent dispersion relation that are normalized by 
$H_0$, $\rho_{00}$, and $t_\mathrm{ff}$.
The result can be applicable to a wide range of parameters simply by using 
the scaling relation on ${\cal M}_0$.

\subsection{Results}
\begin{figure}[htpb]
        \begin{center}
               \includegraphics[width=8cm]{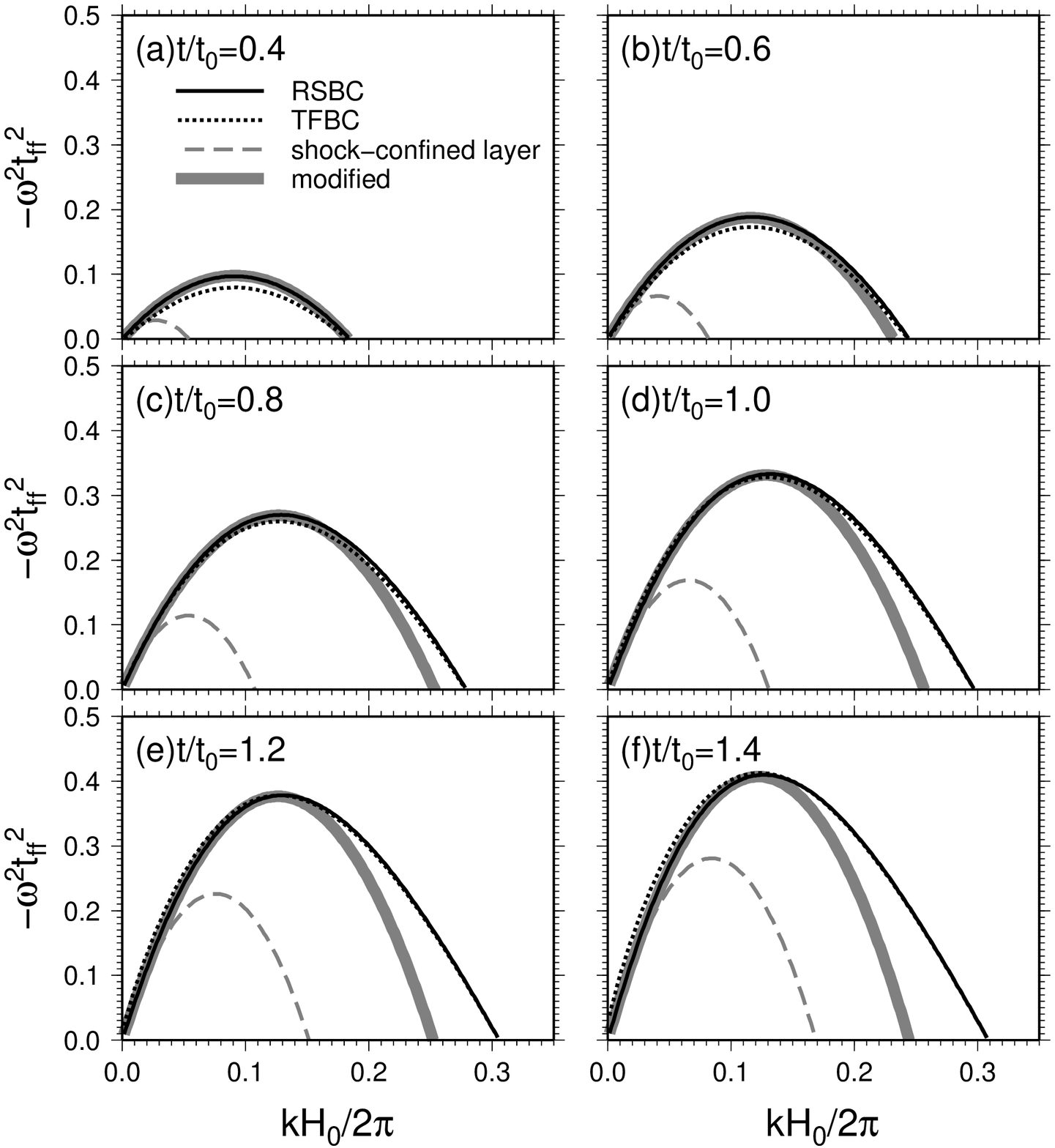}
        \end{center}
\caption{Dispersion relations 
derived from our linear analysis using RSBC (the solid lines) and 
TFBC (the dotted lines). 
For comparison, the dispersion relation of shock-confined layer 
(Equation (\ref{disp shock})) is plotted by the dashed gray lines.
The thick gray lines represent modified dispersion relation (Equation (\ref{disp mod})).
The abscissa and ordinate axes indicate the wavenumber $kH_0/2\pi$ and 
the growth rate normalized by 
$t_\mathrm{ff}=1/\sqrt{2\pi G \rho_{00}}$, respectively.
}
\label{fig:disp}
\end{figure}
\begin{figure}[htpb]
        \begin{center}
               \includegraphics[width=8cm]{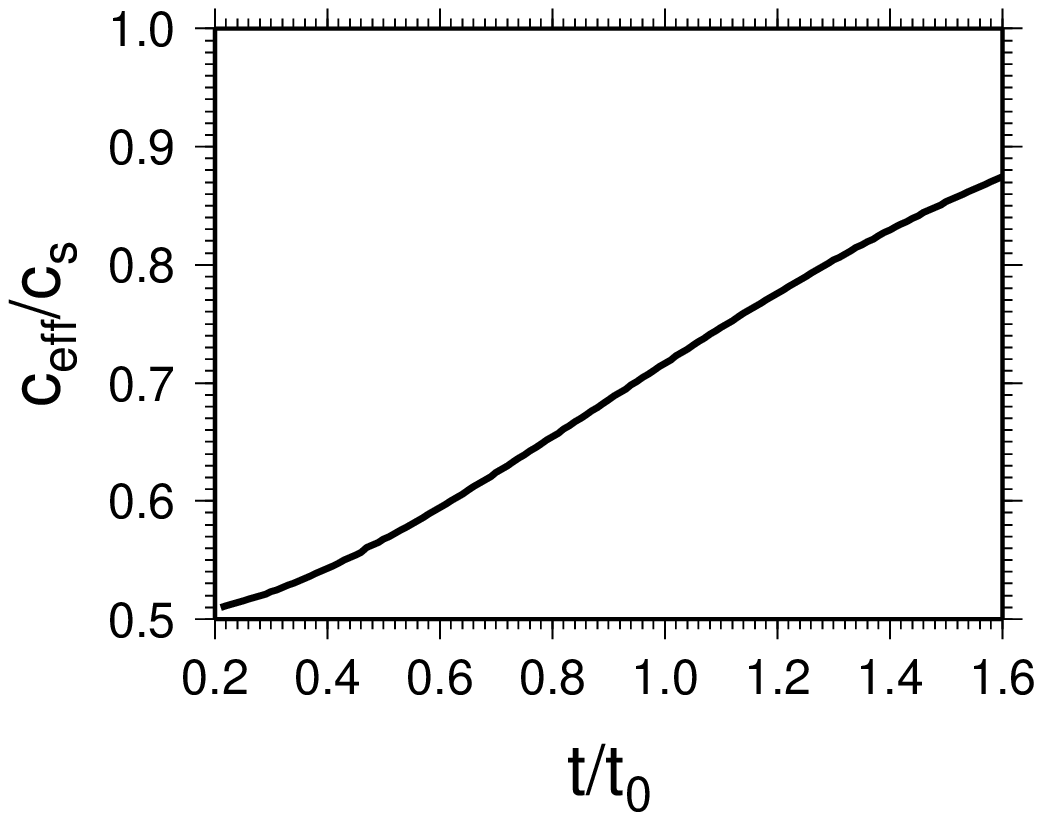}
        \end{center}
   \caption{Time evolution of the ratio of $c_\mathrm{eff}$ to $c_\mathrm{s}$.
}
\label{fig:ratio}
\end{figure}
\begin{figure}[htpb]
        \begin{center}
               \includegraphics[width=8cm]{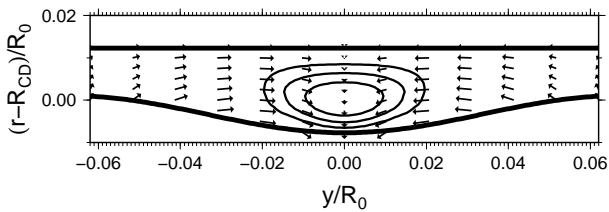}
        \end{center}
\caption{
Cross section of the shell is plotted using eigenfunctions.
The corresponding time and angular wavenumber are $t/t_0=1.3$ and $l=52$, respectively.
The contour indicates the density perturbation normalized by $\rho_{00}$.
The vectors represent velocity perturbations.
}
\label{fig:cross}
\end{figure}
At any time, the unperturbed state is given by the procedure in 
Section \ref{sec:unp}.
Perturbation Equations (\ref{eoc per})-(\ref{poi per}) are solved 
as the eigenvalue- and boundary-value problem. 
As a result, the growth rate, $\omega(k,t)$ can be obtained as a function of 
the wavenumber and time. 

First, we present the results of the linear analysis in 
Figure \ref{fig:disp} at various epochs.  
The ordinate and the abscissa axes represent the non-dimensional growth 
rate $\omega t_\mathrm{ff}$ and wavenumber $kH_0/2\pi$.
The solid and the dotted lines indicate the results of the linear analysis using 
RSBC and TFBC, respectively.
We refer the growth rates obtained by using RSBC and TFBC to 
$\omega_\mathrm{RSBC}$ and $\omega_\mathrm{TFBC}$, respectively.
The dependence of the dispersion relation on the parameters $(n_\mathrm{E},\:Q_\mathrm{UV})$ can be
eliminated by using non-dimensional growth rate $\omega t_\mathrm{ff}$
and wavenumber $kH_0$ as shown in Section \ref{sec:scaling disp}.
We have confirmed that the dispersion relation is identical to that with 
other parameter sets of $(n_\mathrm{E},\;Q_\mathrm{UV})$ 
by using the non-dimensional quantities.
Figure \ref{fig:disp} shows that the difference between 
$\omega_\mathrm{RSBC}$ and $\omega_\mathrm{TFBC}$ is negligible
although RSBC and TFBC are physically quite different. 

In this analysis, we do not take into account the evolutionary effects, such as 
the expansion and accretion of the gas.
Therefore, we compare the results of the linear analysis with the dispersion relation of the shock-confined 
layer (Equation (\ref{disp shock})) rather than that of the expanding shell (Equation (\ref{disp elme})).
One can see that the growth rate is larger than the prediction from the shock-confined layer.
As shown in Section \ref{sec:influence}, this difference comes from the boundary effect of the CD.
Therefore, the shell is expected to begin to grow earlier and to fragment more quickly 
than the prediction from \citet{E94} that 
is based on Equation (\ref{disp shock}).

The dispersion relation with CD + SF boundary conditions is expected to lie between 
that with SF + SF (Equation (\ref{disp shock})) and that with CD + CD (Equation (\ref{disp pre})).
Therefore, to approximate the dispersion relation with RSBC 
analytically, we combine Equation (\ref{disp shock}) with Equation (\ref{disp pre}) as follows:
\begin{equation}
        \omega_\mathrm{mod}^2 = c_\mathrm{eff}^2 k^2 - 2\pi G k 
        \Sigma_\mathrm{s},
        \label{disp mod}
\end{equation}
where $c_\mathrm{eff}$ is the effective sound speed,
\begin{equation}
        c_\mathrm{eff} = \sqrt{ A 2\pi G \Sigma_\mathrm{s} L_\mathrm{eff} 
        + \left( \frac{c_\mathrm{s}}{2} \right)^2},
        \label{soueff}
\end{equation}
where $A$ is a parameter and $L_\mathrm{eff}=\Sigma_\mathrm{s}/\rho_{00}$ is the effective thickness 
that approaches the actual thickness $L_\mathrm{s}$ for small $\Sigma_\mathrm{s}$ and 
$2H_0$ for large $\Sigma_\mathrm{s}$.
The first and the second terms inside the square root correspond to the effect of the CD and the SF boundary 
conditions, respectively.
Here, we choose the parameter $A$  by the condition where
the maximum value of $\omega_\mathrm{mod}$ coincides with
that of $\omega_\mathrm{RSBC}$.
As a result, it is found that a single value of $A=0.39$ shows good 
agreement in growth rates between the modified dispersion relation and 
the detailed linear analysis.
The modified dispersion relations in Equation (\ref{disp mod}) are plotted by the thick gray 
lines in Figure \ref{fig:disp}. 
In Figure \ref{fig:disp}, 
one can see that 
$\omega_\mathrm{mod}$ well describes $\omega_\mathrm{RSBC}$
for $k<k_\mathrm{max}$ all the time.
This suggests that  simply Equation (\ref{disp mod}) can 
describe the most unstable mode obtained by the detailed linear analysis all the time.
Figure \ref{fig:ratio} shows the time evolution of the effective sound speed.
In the early phase, $c_\mathrm{eff}$ is about $0.5c_\mathrm{s}$, suggesting that 
the effect of the CD diminishes the effective sound speed $c_\mathrm{eff}$ by 
half in Equation (\ref{disp shock}).
As the shell expands, $c_\mathrm{eff}$ increases.

The effect of asymmetric density is seen in Figure \ref{fig:disp} where it is found that 
$\omega_\mathrm{mod}$ deviates from $\omega_\mathrm{RSBC}$ for $k>k_\mathrm{max}$
in the gravity-dominated phase ($t/t_0>0.5$).
This is because $\omega_\mathrm{mod}^2$ connects with the P mode $\propto k^2$ 
while $\omega_\mathrm{RSBC}^2$ connects with the SG mode $\propto k$ 
as shown in Section \ref{sec:asym}.

The predicted cross section  of the shell by the linear analysis 
for $(Q_\mathrm{UV}=10^{48.78}$ $\mathrm{s^{-1}}$, $
n_\mathrm{E}=10^3$ $\mathrm{cm}^{-3})$
is shown in Figure \ref{fig:cross} by using the eigenfunctions.
The corresponding time is $t/t_0=1.3$ and the angular wavenumber is $l=52$.
In Figure \ref{fig:cross}, the gas tends to accumulate onto the peak 
only through the upper half region $r>R_\mathrm{c}$.
This property of the flow can be seen from the direction of arrows
in Figure \ref{fig:cross}. Actually, in the upper half region, 
we find $R_\mathrm{SF}-R_\mathrm{c}=1.05H_0>H_0$ that
represents that the gas can collapse to the peak because the sound wave
cannot travel from $R_\mathrm{c}$ to $R_\mathrm{SF}$ within 
the free fall time.
On the other hand, in the region of bottom half ($r<R_\mathrm{c}$),
we find that $R_\mathrm{c}-R_\mathrm{CD}=0.285H_0<H_0$.
This indicates that the gas in $r<R_\mathrm{c}$ cannot collapse to the peak
because the sound wave
can travel from $R_\mathrm{c}$ to $R_\mathrm{CD}$ 
many times within the free fall time. Thus, the pressure 
gradient prevents the compression of gas in the region $r<R_\mathrm{c}$.
However, the GI can proceed even in $r<R_\mathrm{c}$ through 
the deformation of the CD that
makes the gravitational potential deeper.
Therefore, the features of GI in the region $r>R_\mathrm{c}$ and $r<R_\mathrm{c}$ 
have the properties of the ``compressible mode'' and 
``incompressible mode'', respectively.

\section{Discussion}\label{sec:discuss}

The gravitational fragmentation of expanding shells 
confined from both sides by the CD was investigated by \citet{Detal09} numerically and by 
\citet{WDPW10} using analytical approximations.
They assumed that the thermal pressure on both sides is the same and temporally constant.
Therefore, the density peak is always around the mid-plane of the shell, and 
the density profile is almost symmetric.
In their calculation, the column density decreases with time because the shell expands keeping the mass fixed.
Therefore, the pressures at the boundaries approach to the peak pressure.
They found that the confining pressure accelerates fragmentation in the later phase, and  
described this effect as ``pressure-assisted'' gravitational fragmentation. 
This mode is the same as the incompressible mode in this paper. 
\citet{WDPW10} established a semi-analytic linear analysis that
explains results of \citet{Detal09}.

\begin{figure}[htpb]
        \begin{center}
               \includegraphics[width=8cm]{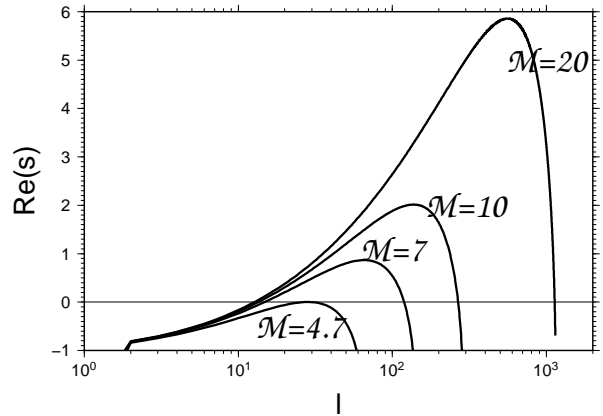}
        \end{center}
\caption{
Growth rate of the VI when the shell expands as $\propto t^{4/7}$.
Each line corresponds to ${\cal M}=4.7$, 10, 7, and 4.7.
}
\label{fig:vishniac}
\end{figure}

\begin{figure}[htpb]
        \begin{center}
               \includegraphics[width=8cm]{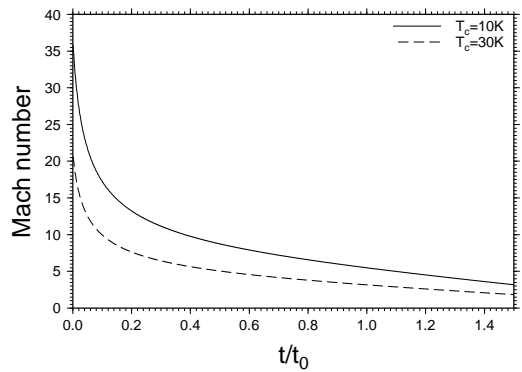}
        \end{center}
\caption{
The Mach number of the shell for $Q_\mathrm{UV}=10^{49}\:s^{\mathrm{-1}}$ and 
$n_\mathrm{E}=10^3$ cm$^{-3}$.
The solid and the dashed lines indicate the case with $T_\mathrm{c}$=10 K and 
30 K, respectively.
}
\label{fig:v}
\end{figure}

The linear analysis in this paper cannot describe the VI 
correctly since the 
approximate shock boundary conditions are imposed in Section \ref{sec:shell}.
The original analysis by \citet{V83} did not find the finite scale 
most unstable mode because the thickness of the shell is neglected.
\citet{VR89} derived a simple analytic dispersion relation of the 
VI for a decelerating isothermal spherical shock wave taking into account 
the effect of the thickness \citep[also see][]{RV87}.
Although, their analysis did not include the self-gravity,
here, we use their dispersion relation 
(see Equations 19(a) and (b) in their paper) 
to estimate the effect of the VI.
Their dispersion relation depends on the Mach number 
${\cal M}$ of the shell and the expansion law.
For the case with the expanding HII regions, the shell expands as 
$\propto t^{4/7}$ if the self-gravity is neglected.
In this case, the perturbation grows not exponentially but in a 
power-law $\propto t^s$, where $s$ characterizes the growth rate. 
Figure \ref{fig:vishniac} shows the real part of $s$ as a function of the angular 
wavenumber $l$.
One can see that the maximum growth rate $\mathrm{Re}(s)$ increases with ${\cal M}$.
The angular scale of the most unstable mode is smaller for larger $\cal M$.
We find that the unstable mode exists only for ${\cal M}\ge4.7$.
To see the typical value of the Mach number,
we consider the expanding shell around the 41$M_{\odot}$ star that is embedded by the 
uniform ambient gas of $n_\mathrm{E}=10^3$ cm$^{-3}$.
Figure \ref{fig:v} shows the Mach number of the shell for $T_\mathrm{c}$ = 10 K (the solid 
line) and 30 K (the dashed line).
In the early phase when the self-gravity is not 
important ($t/t_0<0.5$), 
since the Mach number is as large as several tens, Re$(s)$ is large.
The small scale perturbation with $l=10^2\sim10^3$ quickly grows and saturates in the 
nonlinear stage \citep{MN93}. 
On the other hand, in the later phase (the self-gravity-dominated phase,
$t/t_0>0.5$), 
the Mach number is as low as
$5-10$ as shown in Figure \ref{fig:v}. In this phase, Re$(s)\sim 1$ from
Figure \ref{fig:vishniac}. 
This means that the growth rate of the perturbations is comparable to the 
expansion rate $\propto t^{4/7}$.
Therefore, in the self-gravity-dominated phase, the VI is not expected 
to be important.
The influence of VI on the GI is expected to be only the increase of 
the initial amplitude of perturbations for the GI.

\section{Summary}\label{sec:summary}
In this paper, we have performed linear perturbation analysis 
of decelerating shells created by the expansion of HII regions.
We summarize our results as follows:

\begin{enumerate}
   \item We develop a semi-analytic method for describing the density profile 
         in the shell.
         The time evolution of the density profile of the expanding shell can 
         be divided into three phases, deceleration-dominated, intermediate, and 
         self-gravity-dominated phase.  In the deceleration-dominated phase, 
         the density peak is in SF by the inertia force owing to 
         the deceleration. As the shell mass increases and the self-gravity becomes important, 
         the density peak is inside the shell, but it is closer to the SF than the CD 
         in the intermediate phase. 
         In the self-gravity-dominated phase, the shell becomes massive and the density peak is closer to the CD
         than the SF.
         The evolution is confirmed by 1D hydrodynamical simulation.

   \item We show detailed structures of dispersion relation in the asymmetric layer subjected to a constant
         deceleration both of unstable and stable modes by imposing the CD
         boundary condition from/at both sides.
         \begin{itemize}
                 \item We discover the mode exchange between the compressible 
                       and surface-gravity modes in the stable regime.
                 \item In a situation where the distance from one surface $z_1$ to the density peak $z=0$ 
                       is smaller than the scale height 
                       of the self-gravity $H_0$ and the distance from the other surface $z_2$ to $z=0$
                       larger than $H_0$, the nature of the GI is quite different from the symmetric case with 
                       the same column density and the peak density.
                       The eigenfunction in the region $0<z<z_2$ is approximately
                       the compressible mode. 
                       On the other hand, the eigenfunction in the region $z_1<z<0$ 
                       is approximately the incompressible mode.
                       Moreover, the growth rate is enhanced 
                       compared with symmetric cases through cooperation with the Rayleigh-Taylor instability.
         \end{itemize}

   \item We investigate linear stability of expanding shells driven by HII regions taking into account
         the shock-like boundary condition on the leading surface, the CD
         boundary condition on the trailing surface, and the asymmetric density profile obtained by 
         the semi-analytic method.
         \begin{itemize}
               \item The shell is expected to grow earlier than the prediction of previous studies 
                       \citep{E94,Wetal94b} that are based on the dispersion relation of the shock-confined 
                      layer.
               \item In the self-gravity-dominated phase, since 
                      the density peak is closer to the CD than the SF, 
                      the CD is expected to deform significantly.
         \end{itemize}

\end{enumerate}

These results provide useful knowledge for the analysis of more detailed 
nonlinear numerical simulations that is the scope of our next paper \citep{IIT11}.

\acknowledgments
We thank the referee for many constructive comments that improve our paper significantly.
This work was supported by Grants-in-Aid for Scientific Research
from the MEXT of Japan (K.I.:22864006; S.I.:18540238 and 16077202), 
and Research Fellowship from JSPS (K.I.:21-1979). 
This work was based on the results of the companion paper \citep{IIT11} where 
numerical computations carried out on Cray XT4 at the CfCA of 
Numerical Astronomical Observatory of Japan.
The page charge of this paper is supported by CfCA


\end{document}